\documentstyle[12pt]{article}
 
\begin{document}
\noindent \mbox{\hspace*{58ex}} OCU-PHYS-165 \\ 
\mbox{\hspace*{59ex}} September 1997 

\vspace*{10mm}

\begin{center}
{\Large {\bf 
Finite-temperature reaction-rate formula: Finite volume system, 
detailed balance, $T \to 0$ limit, and cutting rules}} 
\end{center}

\hspace*{3ex}

\hspace*{3ex}

\hspace*{3ex}

\begin{center} 
{\large {\sc A. Ni\'{e}gawa}\footnote{ 
E-mail: niegawa@sci.osaka-cu.ac.jp}

{\normalsize\em Department of Physics, Osaka City University } \\ 
{\normalsize\em Sumiyoshi-ku, Osaka 558, Japan} } \\
\end{center} 

\hspace*{2ex}

\hspace*{2ex}

\hspace*{2ex}

\hspace*{2ex}
\begin{center} 
{\large {\bf Abstract}} \\ 
\end{center} 
\begin{quotation}
A complete derivation, from first principles, of the reaction-rate 
formula for a generic process taking place in a heat bath of finite 
volume is given. It is shown that the formula involves no 
finite-volume correction. Through perturbative diagrammatic analysis 
of the resultant formula, the detailed-balance formula is derived. 
The zero-temperature limit of the formula is discussed. Thermal 
cutting rules, which are introduced in previous work, are compared 
with those introduced by other authors. 
\end{quotation}
\newpage
\setcounter{equation}{0}
\setcounter{section}{0}
\def\theequation{\mbox{\arabic{section}.\arabic{equation}}}
\setcounter{equation}{0}
\setcounter{section}{0}
\def\theequation{\mbox{\arabic{section}.\arabic{equation}}}
\section{Introduction}
Ultrarelativistic heavy-ion-collision experiments at CERN and RHIC 
lead us to entertain a hope of reviving quark-gluon plasma (QGP) in 
the present day. As promising observables of the QGP formation, 
rates of various reactions taking place in a QGP (heat bath) have 
been computed by many authors. Almost all of them, however, 
concentrated on the analyses of particle production from a QGP or 
the decay rate of a particle in a QGP, whose computational method 
has long been known \cite{wel}. 

Since then, through analyses from first principles, a calculational 
scheme of the rate of a generic thermal reaction has been proposed 
\cite{nie,nie-1,nie-2,nie1,nie-tak,kus}. The resultant reaction-rate 
formula is written in terms of the Keldish variant of the real-time 
formalism (RTF) of thermal field theory \cite{l-w}. However, 
complete analysis of classes of diagrams,  which leads to diagrams 
in RTF including thermal propagators with $n$ $(\geq 2)$ thermal 
self-energy insertion, is still lacking. Ref. \cite{nie-1} is the 
only work that discusses such classes of diagrams in scalar field 
theory. In the course of deduction \cite{nie-1} of such diagrams, 
there comes about an involved series, for which an identity is 
assumed. As for \cite{nie-2}, where fermion fields are dealt with, 
the set of diagrams under consideration is not analyzed. This is 
also the case\footnote{In fact, in \cite{kus}, an $n$ ($\geq 2$) 
thermal self-energy-inserted propagator is not deduced from the 
starting formula but is assumed at the start to have the correct 
form in RTF (cf. Eq. (17) in \cite{kus}).} for \cite{kus}. 
Incidentally, the thermal self-energy part in itself and the one 
thermal self-energy-inserted propagator are deduced in 
\cite{nie,nie-1,nie-2,j-l}. 

The principal purpose of this paper is to present a complete 
derivation of the thermal reaction-rate formula (Secs.~II~-~V). 

There has been confusion regarding the issue of finite-volume 
corrections to the standard thermal perturbation theory. (Why and 
how has the confusion arisen is described historically in \cite{eva} 
with relevant references.) By employing a cubic system with periodic 
boundary condition, it has been shown in \cite{eva} that {\em 
thermal expectation values} of normal-ordered products of field 
operators can be chosen to be zero and there is no finite-volume 
correction on thermal amplitudes. It should be stressed that this 
statement is the statement within the RTF. The statement does not 
tell us whether or not the thermal reaction-rate formula deduced 
from first principles is free from finite-volume corrections. We 
shall derive in Secs.~II~-~V the thermal reaction-rate formula for 
the finite-volume system and explicitly see that there is no 
finite-volume correction. 

It should be emphasized that the absence of finite-volume 
corrections here as well as in \cite{eva} is of rather academic 
since a cubic system with periodic boundary condition is taken.  For 
physical finite-volume system, there are \cite{nie-tak} two sources 
of entering the finite-volume effects on the thermal perturbation 
theory constructed on the basis of (grand) canonical ensemble. The 
one comes from the physically sensible boundary condition on the 
single-particle wave function. The other comes from taking the 
physically sensible ensemble. For the case of nonequilibrium case, 
such as expanding QGP, the situation is of course much more 
involved. 

In Sec.~VI, through diagrammatic analysis for the re\-ac\-tion-rate 
formula, we derive the detailed-balance formula. In Sec.~VII, we 
analyze the zero-temperature limit of the reaction-rate formula and 
reproduce a variant of the Cutkosky rules \cite{cut}. 

At zero temperature, the cutting (Cutkosky) rules \cite{cut} are the 
powerful device to investigate the imaginary or absorptive part of 
a scattering amplitude and a reaction rate like a scattering cross 
section. Then, it is natural to infer that a finite-temperature 
extensions of the cutting rules (thermal cutting rules) also plays 
an important role in thermal field theory. 

Previously, several authors 
\cite{kob,nie,nie-1,nie-2,nie1,nie-tak,kus,j-l,kob1,jeo,bed,gel} 
have discussed 
thermal cutting rules.\footnote{Relationship between a thermal 
self-energy part (in im\-ag\-i\-na\-ry-time formalism) and a rate 
of decay 
(production) of a particle in (from) a heat bath was clarified in 
\cite{wel}, from which the cutting rules as applied to the 
self-energy part can be read off.} However, because of the fact that 
the generalization of the notion of \lq\lq cutting'' in vacuum 
theory to the case of thermal field theory is not unique, the terms 
\lq\lq cutting'' and \lq\lq (un)cuttable'' are endowed with 
different meanings in 
\cite{kob,nie,nie-1,nie-2,nie1,nie-tak,kus,j-l,kob1,jeo,bed,gel}, 
which causes recent controversy. With this circumstances in mind, we 
pigeonhole different definitions of thermal cutting rules 
(Sec.~VIII). 
\setcounter{equation}{0}
\setcounter{section}{1}
\section{Preliminary} 
\def\theequation{\mbox{\arabic{section}.\arabic{equation}}}
We consider a heat-bath system of temperature $T$, composed of the 
fields $\phi^{(\alpha)}$, with $\alpha$ labeling collectively a 
field type and its internal degree of freedom. 
We assume $T >> m$ and ignore $m$ (hot plasma). 
The system is inside 
a cube with volume $V = L^3$. Employing the periodic boundary 
conditions, we label the single-particle basis by its momentum 
${\bf p_{\bf k}} = 2\pi {\bf k} /L$, $k_j = 0,\pm 1,\pm 2, \cdots, 
\pm \infty \; (j = x, y, z)$. 
    
Physically interesting thermal reactions are of the following 
generic type,
\begin{equation} 
\{ A \} + \mbox{heat bath} \to \{ B \} + \mbox{anything} \, . 
\label{jyo} 
\end{equation} 
Here $\{ A \}$ and $\{ B \}$ designate group of particles, which are 
not thermalized, such as virtual photons and leptons. 
(Generalization to more general process, where among $\{ A \}$ 
and/or $\{ B \}$ are $\phi^{(\alpha)}$'s, is straightforward 
and will be dealt with in Sec.~V.) The reaction rate ${\cal R}$ of 
the thermal process (\ref{jyo}) is expressed \cite{nie,nie-1,nie-2} 
as an statistical average of the transition probability \( W = 
S^{\ast} S \) (with $S$ the $S$-matrix element) of the {\em 
zero-temperature} ($T = 0$) {\em process}, 
\begin{equation} 
\{ A \} + \{ n_{\bf k}^{(\alpha)} \} \to \{ B \} + \{ 
n_{\bf k}^{(\alpha)}{}' \} \, , 
\label{2} 
\end{equation} 
where $ \{ n_{\bf k}^{(\alpha)} \}$ denotes the group of 
$\phi^{(\alpha)}$'s, which consists of the number $n_{\bf 
k}^{(\alpha)}$ of $\phi_{\bf k}^{(\alpha)}$ ($\phi^{(\alpha)}$ 
in a mode ${\bf k}\; )$:
\begin{eqnarray}
{\cal R} & = & {\cal N} / {\cal D} \:, 
\label{R} \\ 
{\cal N} & \equiv & 
\overline{\sum}_{\{ n_{\bf k}^{(\alpha)} \}} 
\; \rho \; 
\overline{\sum}_{\{ n_{\bf k}^{(\alpha) '} \}} 
\frac{W( \mbox{process $(\ref{2})$} )}{2 \pi \delta(0)} \, , 
\label{cal-N} \\ 
{\cal D} & \equiv & 
\overline{\sum}_{\{ n_{\bf k}^{(\alpha)} \}} 
\; \rho \; 
\overline{\sum}_{ \{ n_{\bf k}^{(\alpha) '} \}} 
W_{ 0 } ( \{ n_{\bf k}^{(\alpha)} \} \to 
\{ {n_{\bf k}}^{(\alpha)}{}' \} ) \, , 
\label{cal-D} \\ 
\rho & = & N^{-1} \; 
\mbox{exp} \left( - {\beta} \sum_{\alpha} \sum_{\bf k} 
n_{\bf k}^{(\alpha)} \, p_{\bf k} \right) \, . 
\label{rho} 
\end{eqnarray}

\noindent Here $\beta = 1 / T$, $p_{\bf k} = |{\bf p}_{\bf k}|$, and 
$2 \pi \delta (0) = t_f - t_i \; (\sim \infty)$ is the time interval 
during which the interaction acts. $W_0 = S_0^\ast S_0$, the \lq\lq 
thermal vacuum bubble,'' is the $T = 0$ transition probability of 
the process indicated in Eq.~(\ref{cal-D}), i.e., the reaction among 
the heat-bath particles $\phi^{(\alpha)}$'s alone. Note that the 
perturbation series for ${\cal D}$ starts from 1, 
\begin{equation} 
{\cal D} = 1 + ... \, . 
\label{series} 
\end{equation} 
In Eq.~(\ref{rho}), $N$ is the normalization factor. In Eqs. 
(\ref{cal-N}) and (\ref{cal-D}), $\overline{\sum}$ stands for 
the summation with symmetry factors being respected, and, for a 
bosonic (fermionic) $\phi^{(\alpha)}$, $n_{\bf k}^{(\alpha)}$ runs 
over $0,1,2, \cdots, \infty$ (0 and 1). It is to be noted that $\{ 
A \}$ and $\{ B \}$ in $S$, which we write $\{A, \, B \}_S$, are not 
necessarily involved in one connected part of $S$. This is also the 
case for $\{A, \, B \}_{S^*}$. We assume that, in $W = S^* S$, $\{A, 
\, B \}_S$ and $\{A, \, B \}_{S^*}$ are involved in one connected 
part, which we simply refer to as connected $W$. Then, a connected 
$W$ consists, in general, of two mutually disconnected parts, the 
one includes $\{A, \, B \}_S$ and $\{A, \, B \}_{S^*}$ and the other 
is a group of spectator particles. Generalization to other cases is 
straightforward. Examples of double-cut diagrams \cite{kin} for 
$S^* S$ are depicted in Fig.~1. It should be remarked on 
the form of $\rho$ in Eq.~(\ref{rho}). Let us recall the following 
two facts. On the one hand, the statistical ensemble is defined by 
the density matrix at the very initial time $t_i$ ($\sim - \infty$). 
On the other hand, in constructing perturbative RTF, an adiabatic 
switching off of the interaction is required \cite{n-s,j-l,l-w}. 
Then, the Hamiltonian $H$ in $\rho \equiv N^{- 1} e^{- \beta H}$ 
should be the free Hamiltonian $H_0$, which leads to 
Eq.~(\ref{rho}). 

As will be seen below, diagrammatic analysis shows that ${\cal N}$, 
Eq.~(\ref{cal-N}), takes the form, 
\begin{equation} 
{\cal N} = {\cal N}_{\scriptsize{con}} \, {\cal D} \, , 
\label{hahaha} 
\end{equation} 
where ${\cal N}_{\scriptsize{con}}$ corresponds to a connected 
diagram and ${\cal D}$ is as in Eq.~(\ref{R}). Then ${\cal R} = 
{\cal N}_{\scriptsize{con}}$. 

The $T=$ $0$ $S$-matrix element is obtained through an application 
of the reduction formula. As an illustration, we take a heat-bath 
system of thermal neutral scalars $\phi$'s, and we take 
$\{ A \}$ to be $\{ \Phi ({\bf p}_i) ; i = 1, ..., m \}$ and 
$\{ B \}$ to be $\{ \Phi ({\bf q}_j) ; j = 1, ..., n \}$ with 
$\Phi$ a nonthermalized heavy neutral scalar. 
Assuming a $\Phi$-$\phi$ coupling to be of the form $\Phi\phi^l$, 
we have \cite{nie,nie-1}
\begin{eqnarray}
S & = & \prod_{j=1}^m \left( iK_{P_j, \Phi_j} \right) \prod_{k=1}^n 
\left( iK_{Q_k, \Phi_k}^* \right) \prod_{\bf k} \left[ 
\sum_{i_{\bf k} = 0}^{n_{\bf k} } \sum_{i'_{\bf k} = 
0}^{n'_{\bf k}} \delta(n_{\bf k}-i_{\bf k} \; ; \; n'_{\bf k} - 
{i'_{\bf k}}) \right. \nonumber \\ 
& & \left. \times N^{n_{\bf k} \, n_{\bf k}{}'}_{i_{\bf k} \, 
i'_{\bf k}} \prod_{n'=1}^{i'_{\bf k}} \left(iK_{{\bf k},n'}^* 
\right) \prod_{n = 1}^{i_{\bf k}} \left( iK_{{\bf k},n} \right) 
\langle 0 \mid T \left[ \prod_{n'=1}^{i'_{\bf k}} \phi_{n'} 
\prod_{n=1}^{i_{\bf k}} \phi_n \prod_{j=1}^m \Phi_j \prod_{k=1}^n 
\Phi_k \right] \mid 0 \rangle \right] \, , \nonumber \\ 
\label{S} 
\end{eqnarray}

\noindent where 
\begin{equation} 
N^{n_{\bf k} n'_{\bf k}}_{i_{\bf k} \, i'_{\bf k}} \equiv 
\left\{ \left(
\begin{array}{c}
{n'_{\bf k}} \\ 
{i'_{\bf k}} 
\end{array}  \right)
\left(
\begin{array}{c}
n_{{\bf k}} \\ 
i_{\bf k}
\end{array} \right) 
\frac{1}{i_{\bf k}'! \; i_{\bf k}!} \right\}^{1/2}. 
\label{N-def} 
\end{equation} 
In Eq.~(\ref{S}) $\delta(\cdots \, ; \, \cdots)$ denotes the 
Kronecker's $\delta$-sym\-bol, 
\begin{eqnarray} 
K_{{\bf k},n} \cdots \phi_{n} & \equiv & {1 \over 
{\sqrt{2 \, p_{\bf k} V Z_{\phi}} }} \int \! d^4 x \, 
e^{-i p_{\bf k} \cdot x} \, \Box \, \cdots \phi (x) \, , \nonumber 
\\ 
K_{P_j, \, \Phi_j} \cdots \Phi_j & \equiv & {1 \over 
{\sqrt{2 \, E_j V Z_{\Phi}} }} \int \! d^4 x \, e^{-i P_j \cdot x} 
\nonumber \\ 
& & \times (\Box + M^2) \, \cdots  \Phi_j (x) \, , 
\label{K} 
\end{eqnarray} 
where $E_j = \sqrt{p_j^2 + M^2}$ with $M$ the mass of $\Phi$. $Z$'s 
in Eq.~(\ref{K}) are the wave-function renormalization constants. 
$S_0$ in $W_0 = S_0^* S_0$ is given by a similar expression to 
Eq.~(\ref{S}), where factors related to the $\Phi$ fields are 
deleted. It is to be noted that, in Eq.~(\ref{S}), among 
$n_{\bf k}$ ($n'_{\bf k}$) of ${\phi_{\bf k}}$'s in the initial 
(final) state, $i_{\bf k}$ ($i'_{\bf k}$) of ${\phi_{\bf k}}$'s are 
absorbed in (emitted from) the $i_{\bf k}$ ($i'_{\bf k}$) vertices 
in $S$. Remaining $n_{\bf k} - i_{\bf k}$ ($= n'_{\bf k} - 
i'_{\bf k}$) of ${\phi_{\bf k}}$'s  are merely spectators, which 
reflects only on the statistical factor in $A$ in Eq.~(\ref{rate}) 
below. 

The expression for $S^*$, the complex conjugate of $S$, is obtained 
by taking the complex conjugate of Eq.~(\ref{S}), where we make the 
substitution, 
\[
i_{\bf k} \to j_{\bf k} \, \;\;\;\;\;\; i'_{\bf k} \to 
j'_{\bf k} \, . 
\]
This applies also to the expression for $S_0^*$. 
\setcounter{equation}{0}
\setcounter{section}{2}
\section{Derivation of the reaction-rate formula} 
\def\theequation{\mbox{\arabic{section}.\arabic{equation}}}
In this section, we take self-interacting neutral scalar theory. 
Generalization to the complex-scalar theory is straightforward (cf. 
Sec.~VIII). A comment on gauge theories is made at the end of this 
section. Fermion fields are dealt with in Sec.~IV. 
\subsection{Analysis of non mode-overlapping diagrams, $i_{\bf k} + 
i'_{\bf k} + j_{\bf k} + j'_{\bf k} \leq 2$} 
In this subsection, for completeness, we briefly recapitulate the 
heart of the analysis of \cite{nie,nie-1}. Let us analyze ${\cal N}$ 
in Eq.~(\ref{cal-N}) with $S$ in Eq.~(\ref{S}). 

(a) $\{ i_{\bf k} = i'_{\bf k} = j_{\bf k} = j'_{\bf k} = 0 \}$. 

Let us take a diagram for $W = S^* S$. Let $v_1$ and $v_2$ be the 
vertices inside $S$, which are connected by the propagator 
\begin{equation} 
\frac{1}{V} \int_{- \infty}^\infty \frac{d p_0}{2 \pi} \, 
\frac{i}{P^2 + i 0^+} \, . 
\label{T0} 
\end{equation} 

(b) $\{ i_{\bf k} = i'_{\bf k} = 1, \; j_{\bf k} = j'_{\bf k} = 0 
\}$ and $\{ i_{- \bf k} = i'_{- \bf k} = 1, \; j_{- \bf k} = 
j'_{- \bf k} = 0 \}$. 

We first deal with the case $\{ i_{\bf k} = i'_{\bf k} = 1, \; 
j_{\bf k} = j'_{\bf k} = 0 \}$. We take out 
the diagram for $W = S^* S$, which is obtained from $W$ above as 
follows. Remove the propagator (\ref{T0}), connect $\phi_{n = 1; \, 
{\bf k}}$, Eq.~(\ref{S}), to the vertex $v_1$ in $S$, and connect 
$\phi_{n' = 1; \, {\bf k}}$ to $v_2$. Here $\phi_{n = 1; \, 
{\bf k}}$ [$\phi_{n' = 1; \, {\bf k}}$] designates that, in 
Eq.~(\ref{S}), $i K_{{\bf k}, n = 1}$ [$i K^*_{{\bf k}, n' = 1}$] 
operates on $\phi_{n = 1}$ [$\phi_{n' = 1}$]. We pick out from 
Eq.~(\ref{S}), 
\begin{equation}
N_{i \, i'}^{n \, n'} = N^{n \, n}_{1 \, 1} = n \, . 
\label{N} 
\end{equation}
Here and below, we suppress the suffix \lq\lq ${\bf k}$'', 
whenever no confusion arises. In $S^*$, $N^{n n'}_{j j'} = 
N^{n n}_{0 0} = 1$. Inserting $N^{n n'}_{j j'} N^{n n'}_{i i'} = n$ 
into Eq.~(\ref{cal-N}) with Eq.~(\ref{rho}), we obtain 
\begin{equation}
\langle n \rangle = \frac{1}{e^{\beta p} - 1} \equiv n_B (p) \, . 
\label{n} 
\end{equation} 
Here $n_B (p) = 1 / (e^{\beta p} - 1)$ is the Bose distribution 
function and the angular brackets denotes the statistical average, 
\[ 
\langle \Omega_n \rangle \equiv \frac{\sum_{n = 0}^{\infty} 
e^{- \beta n p} \Omega_n}{\sum_{n = 0}^{\infty} e^{- \beta n p}} 
\, . 
\] 

Then, in ${\cal N}$ in Eq.~(\ref{cal-N}), the portion corresponding 
to Eq.~(\ref{T0}) turns out to 
\begin{equation} 
\frac{1}{2 p V} \, n_B (p) = \frac{1}{V} \int_{- \infty}^\infty 
\frac{d p_0}{2 \pi} \, \theta (p_0) \, 2 \pi \, \delta (P^2) \, n_B 
(p) \, ,  
\label{T+} 
\end{equation} 
where $1 / (2 p V)$ has come from $i K^*_{{\bf k}, \, n' = 1} 
i K_{{\bf k}, \, n = 1}$ in Eq. (\ref{S}) with Eq.~(\ref{K}). It is 
to be noted that $Z_\phi^{- 1/ 2}$ in $K$'s, Eq.~(\ref{K}), may be 
dealt with just as in vacuum theory, so that we ignore $Z_\phi^{- 1 
/ 2}$ throughout this paper. 

$\{ i_{- \bf k} = i'_{- \bf k} = 1, \; j_{- \bf k} = j'_{- \bf k} = 
0 \}$. 

The relative diagram to the above diagram for $W = S^* S$, same as 
above $W$ except that $\phi_{n = 1; \, - {\bf k}}$ ($\phi_{n' = 1; 
\, - {\bf k}}$) is connected to the vertex $v_2$ ($v_1$), yields, in 
place of Eq.~(\ref{T+}), 
\begin{equation}
\frac{1}{V} \int_{- \infty}^\infty \frac{d 
p_0}{2 \pi} \, \theta (- p_0) \, 2 \pi \, \delta (P^2) \, n_B 
(p) \, . 
\label{T-} 
\end{equation} 

Adding Eqs.~(\ref{T0}), (\ref{T+}), and (\ref{T-}), we extract 
\begin{eqnarray} 
\frac{i}{P_{\bf k}^2 + i 0^+} + 2 \pi \, n_B (p_{\bf k}) \, 
\delta (P_{\bf k}^2) & \equiv & i D_{1 1} (P_{\bf k}) \nonumber \\ 
& \equiv & i D_{1 1}^{(0)} (P_{\bf k}) + i D_{1 1}^{(T)} (P_{\bf k}) 
\, . 
\end{eqnarray} 
Here $i D_{1 1}^{(0)}$ and $i D_{1 1}^{(T)}$ stand, respectively, 
for the $T$-independent part (the first term on the left-hand side 
(LHS)) and the $T$-dependent part (the second term) of $i D_{1 1}$. 

(c) $\{ i_{\bf k} = j_{\bf k} = 0, \; i'_{\bf k} = j'_{\bf k} = 1 
\}$ and $\{ i_{- {\bf k}} = j_{- {\bf k}} = 1, \; i'_{- {\bf k}} = 
j'_{- {\bf k}} = 0 \}$. 

In order to extract the contribution of $\{ i_{\bf k} = j_{\bf k} = 
0, \; i'_{\bf k} = j'_{\bf k} = 1 \}$, we take a diagram 
for $W = S^* S$ in ${\cal N}$, Eq.~(\ref{cal-N}), where $\phi_{n' = 
1; \, {\bf k}}$ in $S$ is connected to the vertex $v_1$ in $S$ and 
$\phi_{n' = 1; \, {\bf k}}$ in $S^*$ is connected to the vertex 
$v_2$ in $S^*$. 

We pick out from Eq.~(\ref{S}) and from the form of $S^*$, 
\[
N^{n \, n'}_{i \, i'} \, 
N^{n \, n'}_{j \, j'} = N_{0 1}^{n, \, n + 1} \, N_{0 1}^{n, \, 
n + 1} = n + 1 \, . 
\]
Inserting into Eq.~(\ref{cal-N}) yields 
\begin{equation} 
1 + n \to 1 + n_B (p) \, . 
\label{stat} 
\end{equation} 
Then, in ${\cal N}$ in Eq.~(\ref{cal-N}), the portion under 
consideration takes the form 
\begin{equation}
\frac{1}{V} \int_{- \infty}^\infty \frac{d p_0}{2 \pi} \, \theta 
(p_0) \, 2 \pi \{1 + n_B (p) \} \, \delta (P^2) \, . 
\label{kataware}
\end{equation}

$\{ i_{- {\bf k}} = j_{- {\bf k}} = 1, \; i'_{- {\bf k}} = 
j'_{- {\bf k}} = 0 \}$. 

We consider the relative diagram for $W = S^* S$, which is the same 
as above except that $\phi_{n = 1; \, - {\bf k}}$ in $S$ is 
connected to the vertex $v_1$ and $\phi_{n = 1; \, - {\bf k}}$ in 
$S^*$ is connected to the vertex $v_2$. In a similar manner as 
above, we have 
\begin{equation} 
\frac{1}{V} \int_{- \infty}^\infty \frac{d p_0}{2 \pi} \, \theta 
(- p_0) \, 2 \pi n_B (p) \, \delta (P^2) \, . 
\label{katawa}
\end{equation}
Adding Eqs.~(\ref{kataware}) and (\ref{katawa}), we extract 
\begin{equation}
2 \pi \, [ \theta (p_{{\bf k} 0}) + n_B (p_{\bf k})] \, 
\delta (P_{\bf k}^2) \equiv i D_{2 1} (P_{\bf k}) \, . 
\label{21} 
\end{equation} 

(d) $\{ i_{\bf k} = i'_{\bf k} = j_{\bf k} = j'_{\bf k} = 0 \}$, 
$\{ i_{\bf k} = i'_{\bf k} = 0, \; j_{\bf k} = j'_{\bf k} = 1 \}$, 
and $\{ i_{- {\bf k}} = i'_{- {\bf k}} = 0, \; j_{- {\bf k}} = 
j'_{- {\bf k}} = 1 \}$. 

In a similar manner as in (a) and (b) above, we extract 
\begin{eqnarray}
\frac{- i}{P_{\bf k}^2 - i 0^+} + 2 \pi \, n_B (p_{\bf k}) \, 
\delta (P_{\bf k}^2) & \equiv & i D_{2 2} (P_{\bf k}) 
\label{d22} \\ 
& \equiv & i D_{2 2}^{(0)} (P_{\bf k}) + i D_{2 2}^{(T)} (P_{\bf k}) 
\nonumber \\ 
& = & (i D_{1 1} (P_{\bf k}))^* \nonumber 
\end{eqnarray}

(e) $\{ i_{\bf k} = j_{\bf k} = 1, \; i'_{\bf k} = j'_{\bf k} = 0 
\}$ and $\{ i_{- {\bf k}} = j_{- {\bf k}} = 0, \; i'_{- {\bf k}} = 
j'_{- {\bf k}} = 1 \}$. 

In a similar manner as in (c) above, we extract 
\begin{eqnarray}
2 \pi \left[ \theta (- p_{{\bf k} 0}) + n_B (p_{\bf k}) \right] \, 
\delta (P_{\bf k}^2) & \equiv & i D_{1 2} (P_{\bf k}) \nonumber \\ 
& = & i D_{2 1} (- P_{\bf k}) \, . 
\label{aha} 
\end{eqnarray}

The forms of $D_{i j} (P)$ ($i, j = 1, 2$) defined above are nothing 
but the thermal propagators in the Keldish variant of RTF, which is 
defined on the time path $C$, $- \infty \to + \infty \to - \infty 
\to - \infty-i{\beta}$, in a complex time plane. The above 
derivation shows that the suffix \lq\lq $1$'' of $D_{i j}$ stands 
for the vertex in $S$ and the suffix \lq\lq $2$'' stands for the 
vertex in $S^*$. On the other hand, in RTF, the suffix \lq\lq $1$'' 
stands for physical or type-1 field and \lq\lq $2$'' stands for 
thermal-ghost or type-2 field. 

Let us turn to identify the vertex factors. We take the interaction 
Lagrangian density, 
\begin{equation} 
{\cal L}_{\mbox{\scriptsize{int}}} = g \Phi 
\phi^\ell / \ell ! + \lambda \phi^{\ell'} / \ell' ! \, . 
\label{i-l} 
\end{equation} 
Then, a $\Phi \phi^\ell$ ($\phi^{\ell'}$) vertex in $S$ receives the 
factor $i g$ ($i \lambda$), and then a $\Phi \phi^{\ell}$ 
($\phi^{\ell'}$) vertex in $S^*$ receives the factor $- i g$ ($- i 
\lambda$). This again is in accord with RTF, where $i g$ ($- i g$) 
and $i \lambda$ ($- i \lambda$) are the factors which are associated 
with, in respective order, $\Phi \phi^\ell$- and 
$\phi^{\ell'}$-vertices of type-1 (type-2) fields. 

Repeating the above procedure, we finally obtain 
\begin{eqnarray} 
& & \frac{1}{V} \left( \prod_{j = 1}^n 2 q_j V \right) {\cal R} 
\nonumber \\ 
& & \mbox{\hspace*{4ex}} = \left( \prod_{i = 1}^m \frac{1}{2 p_i V} 
\right)\, A (P_1^{(2)}, ..., P_m^{(2)}, Q_1^{(1)}, ..., Q_n^{(1)}; 
P_1^{(1)}, ..., P_m^{(1)}, Q_1^{(2)}, ..., Q^{(2)}_n) \, . \nonumber 
\\ 
\label{rate} 
\end{eqnarray} 

\noindent Here $A$ represents the {\em thermal amplitude} in the 
Keldish variant of RTF for the forward process, 
\[ 
\sum_{i = 1}^m \Phi_1 (P_i) + \sum_{j = 1}^n \Phi_2 (Q_j) \to 
\sum_{i = 1}^m \Phi_2 (P_i) + \sum_{j = 1}^n \Phi_1 (Q_j) \, , 
\] 
where $\Phi_1$ ($\Phi_2$) is a type-1 (type-2) field. The thermal 
amplitude $A$ is diagrammed in Fig. 2. As we have assumed 
that $W = S^* S$ represents the connected diagram (cf. above after 
Eq.~(\ref{series})), the diagram for $A$ is connected. 

Each loop momentum $P$ in $A$ accompanies 
\begin{equation} 
\frac{1}{V} \sum_{{\bf p}_{\bf k}} \int \frac{d p_0}{2 \pi} \, . 
\label{loop} 
\end{equation} 

In the large $V$ limit the LHS of Eq.~(\ref{rate}) becomes 
\[ 
\frac{1}{V} \left( \prod_{j = 1}^n 2 q_j V \right) {\cal R} \to 
\frac{1}{V} \left( \prod_{j = 1}^n 2 E_j \frac{d}{d {\bf q}_j / 
(2 \pi)^3} \right) {\cal R} 
\] 
and Eq.~(\ref{loop}) becomes 
\begin{equation} 
\frac{1}{V} \sum_{{\bf p}_{\bf k}} \int \frac{d p_0}{2 \pi} \to 
\int \frac{d^{\, 4} P}{(2 \pi)^4} \, . 
\label{cont} 
\end{equation} 
So far, ${\cal D}$ in Eq.~(\ref{cal-D}) does not participate; 
${\cal D} = 1$ (cf. Eq.~(\ref{series})). The role of ${\cal D}$ will 
be discussed below. 
\subsection{Analysis of mode-overlapping diagrams, $i_{\bf k} + 
i'_{\bf k} + j_{\bf k} + j'_{\bf k} \geq 4$} 
Above derivation of the thermal-reaction-rate formula is not 
complete in that we have only considered the cases where 
$i_{\bf k} + i'_{\bf k} + j_{\bf k} + j'_{\bf k} \leq 2$. When 
generalized self-energy parts are involved in $W = S^* S$, 
$i_{\bf k} + i'_{\bf k} + j_{\bf k} + j'_{\bf k}$ $\geq 4$. [We call 
the diagram with $i_{\bf k} + i'_{\bf k} + j_{\bf k} + j'_{\bf k} 
\geq 4$ the mode-overlapping diagram.] As mentioned in Sec.~I, a 
complete analysis of the classes of diagrams that leads to RTF 
diagrams including thermal propagators with $n$ ($\geq 2$) thermal 
self-energy insertions is still lacking. In this subsection, dealing 
with mode-overlapping diagrams, we shall complete the derivation of 
the thermal-reaction-rate formula. We shall show at the same time 
that there is no finite-volume correction to the formula. 

For illustration of the procedure, we start with analyzing the 
diagram (for $W = S^* S$) with $\{ i_{\bf k} = j_{\bf k} = 
i'_{\bf k} = j'_{\bf k} = 1 \}$. Let us focus our attention on 
$\phi$ with mode ${\bf k}$. Both in $S$ and in $S^*$, there are one 
\lq\lq absorber vertex'' ($v_1'$ and $v_2$ in Fig.~3 below) 
and one \lq\lq emitter vertex'' ($v_1$ and $v_2'$ in 
Fig.~3). 

From $S^* S$, pick out the factor, 
\[ 
N^{n \, n'}_{1 \, 1} N^{n \, n'}_{1 \, 1} = n^2 \, , 
\] 
where and below, the suffix \lq\lq ${\bf k}$'' is dropped whenever 
no confusion arises. In ${\cal N}$ in Eq.~(\ref{cal-N}), we have, in 
place of Eq.~(\ref{n}), 
\begin{eqnarray} 
\langle n^2 \rangle & = & 2 n_B^2 + n_B  \nonumber \\ 
& = & n_B (1 + n_B) + n_B^2 \, , 
\label{decomp} 
\end{eqnarray} 
where $n_B \equiv n_B (p)$. 

The first term on the right-hand side (RHS) of Eq. (\ref{decomp}) 
goes to 
\begin{eqnarray} 
& & \{ 2 \pi \theta (p_0) n_B (p) \, \delta (P^2) \} 
\{ 2 \pi \theta (p_0) [1 + n_B (p)] \, \delta (P^2) \} 
\nonumber \\ 
& & \mbox{\hspace*{5ex}} = i D_{1 2}^{(+)} (P) 
\, i D_{2 1}^{(+)} (P) \, , 
\label{rei-1} 
\end{eqnarray} 
where $D^{(\pm)}_{1 2 / 2 1} (P) \equiv \theta (\pm p_0) \, 
D_{1 2 / 2 1} (P)$. The (part of) thermal propagator $i 
D_{1 2}^{(+)} (P)$ [$i D_{2 1}^{(+)} (P)$] is diagrammed in the 
double-cut diagram for $W = S^* S$, Fig.~3~(a), as the 
line that connects the emitter vertex $v_2'$ [$v_1$] with the 
absorber vertex $v_1'$ [$v_2$]. 

The second term of Eq.~(\ref{decomp}) goes to 
\begin{eqnarray} 
& & \left[  2 \pi \theta (p_0) \, n_B (E) \, \delta (P^2) 
\right]^2 \nonumber \\ 
& & \mbox{\hspace*{5ex}} = i D_{1 1}^{(T) \, (+)} (P) \, i 
D_{2 2}^{(T) \, (+)} (P) \, . 
\label{rei-2} 
\end{eqnarray} 
$i D_{1 1}^{(T) (+)} (P)$ [$i D_{2 2}^{(T) (+)} (P)$] is diagrammed 
in the double-cut diagram, Fig.~3~(b), as the line that 
connects the emitter vertex $v_1$ [$v_2'$] with the absorber vertex 
$v_1'$ [$v_2$]. Thus, with obvious notation, $n_B (1 + n_B)$ part in 
Eq.~(\ref{decomp}) \lq\lq supplies'' $(+ +)$ portion of $i D_{1 2} 
(P) \, i D_{2 1} (P)$, Fig.~3~(a), and $n_B^2$ part \lq\lq 
supplies'' $(+ +)$ part of $i D^{(T)}_{1 1} (P)\, i D^{(T)}_{2 2} 
(P)$ in Fig.~3~(b). 

The $(- -)$ portion of $i D_{1 2} (P) \, i D_{2 1} (P)$ emerges from 
$W = S^* S$, which is the same as Fig.~3 except that 
$\{ i_{- {\bf k}} = j_{- {\bf k}} = i'_{- {\bf k}} = 
j'_{- {\bf k}} = 1 \}$. Now, $v_1$ and $v_2'$ [$v_1'$ and 
$v_2$] are absorber [emitter] vertices. The $(+, -)$ portion comes 
from $W = S^* S$ with $\{ i_{\bf k} = i_{- {\bf k}} = j_{\bf k} 
= j_{- {\bf k}} = 1 \}$. This time, $v_1$ and $v_1'$ [$v_2$ 
and $v_2'$] are absorber [emitter] vertices. The $(- +)$ portion 
comes from $W = S^* S$ with $\{ i'_{\bf k} = i'_{- {\bf k}} = 
j'_{\bf k} = j'_{- {\bf k}} = 1 \}$, where the absorber 
[emitter] vertices are $v_2$ and $v_2'$ [$v_1$ and $v_1'$]. Adding 
all these contributions to the contribution (\ref{rei-1}), we obtain 
Eq.~(\ref{rei-1}) with complete $i D_{1 2} (P) \, i D_{2 1} (P)$. In 
a similar manner, we can find a set of relative diagrams, which, 
together with Eq.~(\ref{rei-2}), yield the complete $i D_{1 1} (P) 
\, i D_{2 2} (P)$. 

All the vertices \lq\lq $v_1$'', \lq\lq $v_1'$'', \lq\lq $v_2$'', 
and \lq\lq $v_2'$'' (cf. Fig.~3) are not necessarily within 
one connected diagram. There is a diagram as depicted, e.g., in 
Fig.~4. Figure 4~(a) [(b)] contains the factor $i 
D_{1 2}^{(+)} \, iD_{2 1}^{(+)}$ [$i D_{1 1}^{(T) (+)} \, 
iD_{2 2}^{(T) (+)}$] in Eq.~(\ref{rei-1}) [Eq.~(\ref{rei-2})]. Let 
us inspect Fig.~4~(a). As stated above after 
Eq.~(\ref{series}), we are considering the case where $\{ A, B \}_S$ 
and $\{ A, B \}_{S^*}$ [$\{ A \} = \{ \Phi (P_i) \}$ and $\{ B \} = 
\{ \Phi (Q_j) \}$] are involved in one connected part of $W = S^* 
S$. Then, all $\Phi$'s are in, e.g., the bottom subdiagram in 
Fig.~4~(a) (and then also in Fig.~4~(b)) and, in 
the middle subdiagram, only constituent particles $\phi$'s of the 
heat bath participate. $i D_{2 1}^{(+)} (P)$ is involved in the 
middle subdiagram, which goes to ${\cal D}$, while $i D_{1 2}^{(+)} 
(P)$ is involved in the bottom subdiagram, which goes to 
${\cal N}_{\scriptsize{con}}$. Thus, Fig.~4~(a) is in 
${\cal N}_{\scriptsize{con}} \, {\cal D}$ with ${\cal D} \neq 1$ in 
Eq.~(\ref{hahaha}) with Eq.~(\ref{series}). As a matter of fact, 
${\cal N}_{\scriptsize{con}}$ here is obtained from $W = S^* S$ with 
$\{ i_{\bf k} = j_{\bf k} = 1, \; i'_{\bf k} = 
j'_{\bf k} = 0 \}$ and ${\cal D}$ is obtained from $W_0 = 
S_0^* S_0$ (cf. Eq.~(\ref{cal-D})) with $\{ i_{\bf k} = 
j_{\bf k} = 0, \; i'_{\bf k} = j'_{\bf k} = 1 \}$. Thus, 
Fig.~4~(a) does contribute to ${\cal R}$ 
in Eq.~(\ref{R}) as ${\cal R}$ $=$ ${\cal N}_{\scriptsize{con}}$, 
which already appears at lower order of perturbation series. As 
above, it is straightforward to find a set of relative diagrams, 
which, together with Fig.~4~(a), yields the complete $i 
D_{1 2} (P) \, i D_{2 1} (P)$. Similarly one can find a set of 
relative diagrams, which, together with Fig.~4~(b), yields 
the complete $i D_{1 1} (P) \, i D_{2 2} (P)$. 

The relevant part of Fig.~4~(b) and its \lq\lq relatives'' 
sits in $A$, Eq.~(\ref{rate}), as a $(1, \, 2)$ component of a 
thermal self-energy-inserted propagator. Thus, $W = S^* S$ with 
$\{ i_{\bf k} = j_{\bf k} = i'_{\bf k} = j'_{\bf k} = 1 \}$ 
together with its \lq\lq relatives'' has turned out to 
take the proper seat in $A$ in Eq.~(\ref{rate}). 

It is straightforward to generalize the above argument to a generic 
diagram for $W = S^* S$. Let us focus our attention on a mode ${\bf 
k}$. We analyze ${\cal N}$ in Eq.~(\ref{cal-N}). Let $\phi_{{\bf 
k}}$ be $\phi$ in the mode ${\bf k}$. In $S$ in Eq.~(\ref{S}), 
$i_{\bf k}$ $\phi_{{\bf k}}$'s in the initial state and 
$i'_{\bf k}$ $\phi_{{\bf k}}$'s in the final state participate 
directly in the reaction. In $S^*$, $j_{\bf k}$ ($j'_{\bf k}$) 
$\phi_{{\bf k}}$'s in the initial (final) state participate 
directly; $i_{\bf k} - i'_{\bf k} = j_{\bf k} - j'_{\bf k} = 
n_{\bf k} - n'_{\bf k}$. In $S$, there are $i_{\bf k}$ 
($i'_{\bf k}$) \lq\lq absorber vertices'' (\lq\lq emitter 
vertices'') and, in $S^*$, there are $j_{\bf k}$ ($j'_{\bf k}$) 
\lq\lq emitter vertices'' (\lq\lq absorber vertices''). [Recall 
that, in the case of Figs.~3 and 4, $v_1'$ and 
$v_2$ are absorber vertices and $v_1$ and $v_2'$ are emitter 
vertices.] 

We pick out from $W = S^* S$, 
\begin{eqnarray} 
N^{n \, n'}_{j \, j'} \, 
N^{n \, n'}_{i \, i'} & = & \frac{n ! \, n' !}{(n - i) ! 
\, (n - j) !} \, 
\frac{1}{i ! \, i' ! \, j ! \, j' !} \nonumber \\ 
& = & \frac{1}{i ! \, i' ! \, j ! \, j' !} 
\prod_{k = 0}^{i' - 1} (n + i' - i - k) \prod_{k = 0}^{j - 1} 
(n - k) \, \, , 
\label{comb} 
\end{eqnarray} 
where and below the suffix \lq\lq ${\bf k}$'' has been dropped. From 
the form for $S$, Eq.~(\ref{S}), we see that the permutation of 
$\phi_{n'}$ ($n' = 1, ..., i'_{\bf k}$) and the permutation of 
$\phi_n$ ($n = 1, ..., i_{\bf k}$) give the same diagram, and then 
$i_{\bf k} ! \, i'_{\bf k} !$ same diagrams emerge. Then $i_{\bf k} 
! \, i'_{\bf k} ! \, j_{\bf k} ! \, j'_{\bf k} !$ same diagrams 
emerge for $W = S^* S$, which {\em eliminates} the first factor on 
the RHS of Eq.~(\ref{comb}). In ${\cal N}$ in Eq.~(\ref{cal-N}), we 
have, in place of Eq.~(\ref{n}), 
\[ 
\langle \prod_{k = 0}^{i' - 1} (n + i' - i - k) \prod_{k = 0}^{j 
- 1} (n - k) \rangle \equiv H^{i, \, i'}_{j, \, j'} \, . 
\] 
Here it is convenient to introduce a generating function of $H^{i, 
\, i'}_{j, \, j'}$, 
\begin{equation} 
f (y, z) \equiv \sum_{n = 0}^{\infty} y^{n + i' - i} \, z^n 
e^{- x n} \;\;\;\;\;\;\;\; (x = \beta p = \beta p_{\bf k}) \, . 
\label{gen} 
\end{equation} 
In fact, from Eq.~(\ref{gen}), we obtain 
\begin{eqnarray} 
H^{i, \, i'}_{j, \, j'} & = & \frac{1}{f} 
\frac{\partial^2 f}{\partial y^{i'} \partial x^{j}} 
\rule[-3.5mm]{.14mm}{9.5mm} \raisebox{-3mm}{\scriptsize{$\; y = z = 
1$}} \, . 
\label{zama} 
\end{eqnarray} 

From Eq.~(\ref{zama}) with Eq.~(\ref{gen}), it can be shown 
that 
\begin{eqnarray} 
H^{i, \, i'}_{j, \, j'} & = & 
\sum_{k = 0}^{\mbox{\scriptsize{min}} (i', \, j')} 
\left( 
\begin{array}{c} 
i' \\ 
k 
\end{array} 
\right) 
\frac{j' ! \, (j + i' - k) !}{(j' - k) !} \{ n_B (x) \}^{j + i' - k} 
\, . 
\label{first} 
\end{eqnarray} 
Since $i - i' = j - j'$, we can readily see that 
$H^{i, \, i'}_{j, \, j'}$, Eq.~(\ref{first}), is symmetric under 
$(i, \, i')$ $\leftrightarrow$ $(j, \, j')$. Then, without loss of 
generality, we assume $i \geq j$. 

In Appendix A, we show that 
\begin{eqnarray} 
H^{i, \, i'}_{j, \, j'} & = & 
\sum_{k = 0}^{\mbox{\scriptsize{min}} (j, \, j')} \frac{i !}{(i 
- j + k) !} \frac{i' ! \, j' !}{(j' - k) !} \left( 
\begin{array}{c} 
j \\ 
k 
\end{array} 
\right) (n_B)^{i + k} ( 1 + n_B )^{j' - k} 
\label{koutou0} \\ 
& = & 
\sum_{k = 0}^{\mbox{\scriptsize{min}} (j, \, j')} 
\left\{ C^k_{i, \, j} \, (n_B)^{j - k} \right\} 
\left\{ C^0_{i', \, i - j + k} \, (n_B)^{i - j + k} \right\} 
\nonumber \\ 
& & \times \left\{ C^0_{j', \, k} \, (n_B)^k \right\} 
\left\{ C^0_{j' - k, \, j' - k} \, (1 + n_B )^{j' - k} 
\right\} \, . 
\label{koutou1} 
\end{eqnarray} 

\noindent Here $n_B \equiv n_B (p)$ and 
\[ 
C^k_{i, \, j} \equiv \frac{i!}{(i - j + k)!} 
\left( \begin{array}{c} 
j \\ 
k 
\end{array} 
\right) \, . 
\] 
In Eq.~(\ref{koutou1}), the factor $C_{i, \, j}^k$ may be identified 
to the number of ways of connecting $j - k$ (out of $j$) emitter 
vertices in $S^*$ to $i$ absorber vertices in $S$, the factor 
$C_{i', \, i - j + k}^0$ to the number of ways of connecting $i - j 
+ k$ absorber vertices in $S$ to $i'$ emitter vertices in $S$, the 
factor $C_{j', \, k}^0$ to the number of ways of connecting $k$ 
emitter vertices in $S^*$ to $j'$ absorber vertices in $S^*$, and 
the factor $C_{j' - k, \, j' - k}^0$ to the number of ways of 
connecting $j' - k$ absorber vertices in $S^*$ to $i' - (i - j + k)$ 
$ = $ $j' - k$ emitter vertices in $S$. Then, in ${\cal R}$ in 
Eqs.~(\ref{R}), we have, in place of Eqs.~(\ref{rei-1}) and 
(\ref{rei-2}), 
\begin{eqnarray} 
& & \sum_{k = 0}^{\mbox{\scriptsize{min}} (j, \, j')} 
\left[ C^k_{i, \, j} \{ i D_{1 2}^{(+)} (p) \}^{j - k} \right] 
\left[ C^0_{i', \, i - j + k} \{ i D_{1 1}^{(T)(+)} (p) \}^{i - 
j + k} \right] 
\nonumber \\ 
& & \mbox{\hspace*{4ex}} \times \left[ C^0_{j', \, k} \{ i 
D_{2 2}^{(T)(+)} (p) \}^k \right] 
\left[ C^0_{j' - k, \, j' - k} \{ i D_{2 1}^{(+)} (p) \}^{j' - k} 
\right] \, . 
\label{yare} 
\end{eqnarray} 

\noindent This is just a portion of \lq\lq right'' thermal amplitude 
in RTF. Just as in the simple case, $\{ i_{\bf k} = j_{\bf k} = 
i'_{\bf k} = j'_{\bf k} = 1 \}$, analyzed above, we can find a set 
of relative diagrams for $W = S^* S$, which, together with 
Eq.~(\ref{yare}), leads to Eq.~(\ref{yare}) with complete $D$'s. 
Among the diagrams that accompany Eq.~(\ref{yare}) with complete 
$D$'s, are disconnected ones like Fig.~4~(a). Such 
diagrams belong to ${\cal N} = {\cal N}_{\scriptsize{con}} \, 
{\cal D}$ with ${\cal D} \neq 1$ (cf. Eq.~(\ref{hahaha})), and then 
do contribute to ${\cal R}$ in Eq.~(\ref{R}) as ${\cal R}$ $ = $ 
${\cal N}_{\scriptsize{con}}$. Connected diagrams that accompany 
Eq.~(\ref{yare}) with complete $D$'s take the proper seat in $A$ in 
Eq.~(\ref{rate}). 

Conversely, for any diagram for $A$ in Eq.~(\ref{rate}), through the 
analysis running in the opposite direction, one can identify a set 
of diagrams for $W = S^* S$. The analysis made above is so general 
that no additional comment is necessary on the diagrams that leads 
to $A$, Eq.~(\ref{rate}), which includes thermal propagator(s) with 
$n$ $(\geq 2)$ thermal self-energy insertion. 

This completes the derivation of the formula (\ref{rate}) for the 
rate of a generic thermal reaction taking place in a heat bath of 
finite volume. Keeping in mind a suitable normalization for incident 
fluxes of $\Phi$'s, the formula (\ref{rate}) \lq\lq smoothly'' goes 
to the formula for the infinite-volume ($V = \infty$) system (cf. 
Eq.~(\ref{cont})) in the sense that there do not exist extra 
contributions in Eq.~(\ref{rate}) with $V < \infty$, which disappear 
in the limit $V \to \infty$. Thus, there is no finite-volume 
correction to the thermal reaction-rate formula (\ref{rate}). 

Here we make a comment on gauge theories. Choosing a physical gauge 
like Coulomb gauge, the gauge boson may be dealt with in a similar 
manner to the above scalar-field case. When we adopt a covariant 
gauge, Faddeev-Popov (FP) ghost field comes on the stage. The first 
summations in Eqs.~(\ref{cal-N}) and (\ref{cal-D}) are carried out 
over the modes of physical degrees of freedom. This can be 
implemented by inserting the projection operator ${\cal P}$ onto the 
physical space on the left side of $\rho$ in Eqs.~(\ref{cal-N}) and 
(\ref{cal-D}) and sum is taken over $\{ n^{(\alpha)}_{\bf k} \}$ for 
all, unphysical as well as physical, modes $\alpha$'s. As far as the 
ensemble average of physical quantities like reaction rate are 
concerned, all the role of ${\cal P}$ is to make \cite{HK} the 
antiperiodic boundary condition for FP-ghost field the periodic one, 
$\phi_{\scriptsize{FP}} (t - i \beta, {\bf x})$ $ = $ 
$\phi_{\scriptsize{FP}} (t, {\bf x})$, so that the bare FP-ghost 
propagator is the same in form to the scalar propagator. Keeping 
this fact in mind, we can deduce Eq.~(\ref{rate}), where $A$ is 
evaluated using standard gauge-field and FP-ghost thermal 
propagators in the covariant gauge. 
\setcounter{equation}{0}
\setcounter{section}{3}
\section{The Dirac fermion} 
\def\theequation{\mbox{\arabic{section}.\arabic{equation}}}
We study the case of Dirac fermion. The expression for $S$ in 
Eq.~(\ref{S}) with Eqs.~(\ref{N-def}) and (\ref{K}) is changed 
accordingly. Let $n_{\bf k}^{(\sigma)}$ [$\overline{n}_{\bf 
k}^{(\sigma)}$] ($\sigma = \pm$) be the number of mode-${\bf k}$ 
fermion [anti fermion] with helicity $\sigma$. The combinatorial 
factor $N^{n_{\bf k} n'_{\bf k}}_{i_{\bf k} i'_{\bf k}}$ in 
Eq.~(\ref{S}) is changed to 
\begin{eqnarray} 
N_f & = & \prod_{\sigma = \pm} \left( 
N^{n_{\bf k}^{(\sigma)} n_{\bf k}^{(\sigma) '}}_{i_{\bf 
k}^{(\sigma)} i_{\bf k}^{(\sigma) '}} \, N^{\overline{n}_{\bf 
k}^{(\sigma)} \overline{n}_{\bf k}^{(\sigma) '}}_{\overline{i}_{\bf 
k}^{(\sigma)} \overline{i}_{\bf k}^{(\sigma) '}} \right) \nonumber 
\\ 
& \equiv & \prod_{\sigma = \pm} \left[ \left( 
\begin{array}{c} 
n_{\bf k}^{(\sigma) '} \\ 
i_{\bf k}^{(\sigma) '} 
\end{array} 
\right) 
\left( 
\begin{array}{c} 
n_{\bf k}^{(\sigma)} \\ 
i_{\bf k}^{(\sigma)} 
\end{array} 
\right) 
\left( 
\begin{array}{c} 
\overline{n}_{\bf k}^{(\sigma) '} \\ 
\overline{i}_{\bf k}^{(\sigma) '} 
\end{array} 
\right) 
\left( 
\begin{array}{c} 
\overline{n}_{\bf k}^{(\sigma)} \\ 
\overline{i}_{\bf k}^{(\sigma)} 
\end{array} 
\right) \right] \, , 
\label{fern} 
\end{eqnarray} 
where $n_{\bf k}^{(\sigma)} - i_{\bf k}^{(\sigma)} = n_{\bf 
k}^{(\sigma)'} - i_{\bf k}^{(\sigma)'}$ and $\overline{n}_{\bf 
k}^{(\sigma)} - \overline{i}_{\bf k}^{(\sigma)} = \overline{n}_{\bf 
k}^{(\sigma)'} - \overline{i}_{\bf k}^{(\sigma)'}$. In 
Eqs.~(\ref{cal-N}) and (\ref{cal-D}), the summations on $n_{\bf 
k}^{(\sigma)}$, $n_{\bf k}^{(\sigma) '}$, $\overline{n}_{\bf 
k}^{(\sigma)}$, and $\overline{n}_{\bf k}^{(\sigma) '}$ are taken 
over 0 and 1. We assume that the interaction Lagrangian is bilinear 
in fermion fields, which include fermion fields constituting the 
heat bath and possibly nonthermalized heavy fermion fields, the 
counterpart of $\Phi$'s in Eq.~(\ref{S}). 
\subsection{Analysis of non mode-overlapping diagrams}
We proceed as in Sec.~III~A using the same notation. 

(a) $\{ i_{\bf k}^{(\sigma)} = i_{\bf k}^{(\sigma) '} = j_{\bf 
k}^{(\sigma)} = j_{\bf k}^{(\sigma) '} = \overline{i}_{\bf 
k}^{(\sigma)} = \overline{i}_{\bf k}^{(\sigma) '} = 
\overline{j}_{\bf k}^{(\sigma)} = \overline{j}_{\bf k}^{(\sigma) '} 
= 0 \}$ ($\sigma = \pm$). 

In place of Eq.~(\ref{T0}), we have 
\begin{equation} 
\frac{1}{V} \int_{- \infty}^\infty \frac{d p_0}{2 \pi} \, 
\frac{i {P\kern-0.1em\raise0.3ex\llap{/}\kern0.15em\relax}}{P^2 + i 
0^+} \, , 
\label{T0-f} 
\end{equation} 
which comes from the following contraction in $S$ (cf. 
Eq.~(\ref{S})), 
\begin{eqnarray} 
& & \langle 0 | \, T \left[ \, \cdot \cdot \cdot \, 
\overline{\psi} (x_1) \, \psi (x_1) \, \cdot \cdot \cdot \, 
\overline{\psi} (x_2) \, \psi (x_2) \, \cdot \cdot \cdot \, \right] 
| 0 \rangle \mbox{\hspace{-29.3ex}} \rule[2.3ex]{.08ex}{.6ex} 
\rule[2.9ex]{10.2ex}{.1ex} \rule[2.3ex]{.08ex}{.6ex} 
\label{cont1} \\ 
& & \mbox{\hspace*{5ex}} = i S_F (x_1 - x_2) \langle 0 | T \left[ 
\, \cdot \cdot \cdot \, \overline{\psi} (x_1) \, \cdot \cdot \cdot 
\, \psi (x_2) \, \cdot \cdot \cdot \, \right] | 0 \rangle \, . 
\nonumber 
\end{eqnarray} 
Here $\overline{\psi} \psi$'s in Eq.~(\ref{cont1}) come from the 
interaction Lagrangian ${\cal L}_{\scriptsize{int}}$. 

(b) Fermion mode with $\{ i_{\bf k}^{(\sigma)} = 
i_{\bf k}^{(\sigma) '} = 1, \; j_{\bf k}^{(\sigma)} = 
j_{\bf k}^{(\sigma) '} = 0 \}$ ($\sigma = \pm$) and its relative. 

We consider the positive-helicity ($\sigma = +$) fermion mode with 
$\{ i_{\bf k}^{(+)} = i_{\bf k}^{(+) '} = 1, \; j_{\bf k}^{(+)} = 
j_{\bf k}^{(+) '} = 0 \}$. In place of Eqs.~(\ref{N}) and (\ref{n}), 
we have, in respective order, 
\[ 
N_f = n^2 \, . 
\] 
and 
\[ 
\langle n^2 \rangle = \frac{1}{e^{\beta p} + 1} \equiv n_F (p) 
\, , 
\] 
where 
\[ 
n_F (p) \equiv 1 / (e^{\beta p} + 1) 
\] 
is the 
Fermi-distribution function and $\langle \Omega_n 
\rangle \equiv \sum_{n = 0}^1 e^{- \beta n p} \Omega_n / 
\sum_{n = 0}^1 e^{- \beta n p}$. We note that the contribution 
corresponding to Eq.~(\ref{cont1}) above is (cf. Eq.~(\ref{S})) 
\begin{eqnarray*} 
& & \langle 0 | \, T \left[ \, \cdot \cdot \cdot \, \psi_{n' = 1} 
(y) \, \overline{\psi} (x_1) \, \psi (x_1) \, \cdot \cdot \cdot \, 
\overline{\psi} (x_2) \, \psi (x_2) \, \overline{\psi}_{n = 1} (z) 
\, \cdot \cdot \cdot \, \right] | 0 \rangle 
\mbox{\hspace{-50.3ex}} 
\rule[2.3ex]{.08ex}{.6ex} 
\rule[2.9ex]{25.8ex}{.1ex} 
\rule[2.3ex]{.08ex}{.6ex} 
\mbox{\hspace{-10.9ex}} 
\rule[-2.1ex]{.08ex}{.6ex} 
\rule[-2.1ex]{22.7ex}{.1ex} 
\rule[-2.1ex]{.08ex}{.6ex} 
\nonumber \\ 
& & \mbox{\hspace*{5ex}} = - i S_F (y - x_2) \, i S_F (x_1 - z) 
\langle 0 | \, T \left[ \, \cdot \cdot \cdot \, \overline{\psi} 
(x_1) \, \cdot \cdot \cdot \, \psi (x_2) \, \cdot \cdot 
\cdot \, \right] | 0 \rangle \, . 
\end{eqnarray*} 

\noindent Then, the LHS of Eq.~(\ref{T+}) is replaced by 
\[ 
- \frac{1}{2 p V} \, n_F (p) \, u^{(+)} (P) \, 
\overline{u}^{(+)} (P) \, . 
\] 
Adding the contribution from the negative-helicity fermi\-on mode 
with $\{ i_{\bf k}^{(-)} = i_{\bf k}^{(-) '} = 1, \; j_{\bf k}^{(-)} 
= j_{\bf k}^{(-) '} = 0 \}$, we have 
\begin{eqnarray} 
& & - \frac{1}{2 p V} \, n_F (p) \sum_{\sigma = \pm} u^{(\sigma)} 
(P) \, \overline{u}^{(\sigma)} (P) \nonumber \\ 
& & \mbox{\hspace*{5ex}} = - \frac{1}{V} \int_{- \infty}^\infty 
\frac{d p_0}{2 \pi} \, \theta (p_0) \, 2 \pi \, \delta (P^2) 
n_F (p) \, {P\kern-0.1em\raise0.3ex\llap{/}\kern0.15em\relax} \, . 
\label{D} 
\end{eqnarray} 
Adding further the contribution from the antifermion modes with 
$\{ \overline{i}_{- {\bf k}}^{(\sigma)} = \overline{i}_{- {\bf 
k}}^{(\sigma) '} = 1, \; \overline{j}_{- {\bf k}}^{(\sigma)} = 
\overline{j}_{- {\bf k}}^{(\sigma) '} = 0 \}$ ($\sigma = \pm$) to 
Eqs.~(\ref{T0-f}) and (\ref{D}), we extract 
\begin{eqnarray} 
& & \left[ \frac{i}{P_{\bf k}^2 + i 0^+} - 2 \pi \, n_F (p_{\bf k}) 
\, \delta (P_{\bf k}^2) \right] 
{P\kern-0.1em\raise0.3ex\llap{/}\kern0.15em\relax}_{\bf k} \nonumber 
\\ 
& & \mbox{\hspace*{4ex}} \equiv i S_{1 1} (P_{\bf k}) = i 
S_{1 1}^{(0)} (P_{\bf k}) + i S_{1 1}^{(T)} (P_{\bf k}) \, . 
\end{eqnarray} 

(c) Fermion mode with $\{ i_{\bf k}^{(\sigma)} = 
j_{\bf k}^{(\sigma)} = 0, \; i_{\bf k}^{(\sigma) '} = 
j_{\bf k}^{(\sigma) '} = 1 \}$ ($\sigma = \pm$) and its relative. 

In place of Eq.~(\ref{stat}), we have 
\[ 
1 - n_F (p) \, . 
\] 
Then, Eq.~(\ref{kataware}) is replaced by 
\[ 
\frac{1}{V} \int_{- \infty}^\infty \frac{d p_0}{2 \pi} \, \theta 
(p_0) \, 2 \pi \{ 1 - n_F (p) \} \, \delta (P^2) \, 
{P\kern-0.1em\raise0.3ex\llap{/}\kern0.15em\relax} \, . 
\] 
Adding the contribution from the antifermion mode with 
$\{ \overline{i}_{- {\bf k}}^{(\sigma)} = \overline{j}_{- {\bf 
k}}^{(\sigma)} = 1, \; \overline{i}_{- 
{\bf k}}^{(\sigma) '} = \overline{j}_{- {\bf k}}^{(\sigma) '} = 0 
\}$ ($\sigma = \pm$), we extract 
\[ 
2 \pi [ \theta (p_0) - n_F (p_{\bf k}) ] \delta (P_{\bf k}^2) \, 
{P\kern-0.1em\raise0.3ex\llap{/}\kern0.15em\relax}_{\bf k} 
\equiv i S_{2 1} (P_{\bf k}) \, . 
\] 

(d) Interchanging the roles of $S$ and $S^*$ in (a) and (b) above, 
we obtain, in place of Eq.~(\ref{d22}), 
\begin{eqnarray*} 
& & \left[ \frac{- i}{P_{\bf k}^2 - i 0^+} - 2 \pi \, n_F 
(p_{\bf k}) \, \delta (P_{\bf k}^2) \right] 
{P\kern-0.1em\raise0.3ex\llap{/}\kern0.15em\relax}_{\bf k} \nonumber 
\\ 
& & \mbox{\hspace*{4ex}} \equiv i S_{2 2} (P_{\bf k}) = i 
S_{2 2}^{(0)} (P_{\bf k}) + i S_{2 2}^{(T)} (P_{\bf k}) \, . 
\end{eqnarray*} 

(e) Fermion mode with $\{ i_{\bf k}^{(\sigma)} = 
j_{\bf k}^{(\sigma)} = 1, \, i_{\bf k}^{(\sigma) '} = 
j_{\bf k}^{(\sigma) '} = 0 \}$ ($\sigma = \pm$) and its relative. 

The relevant statistical factor is $n_F (p)$. Let us show that the 
part under consideration turns out to $i S_{1 2} (P_{\bf k})$. 
In place of $p_0 > 0$ portion of Eq.~(\ref{aha}), we have 
$2 \pi n_F (p_{\bf k})\, \delta (P_{\bf k}^2) 
{P\kern-0.1em\raise0.3ex\llap{/}\kern0.15em\relax}_{\bf k}$ 
which seems to be the $p_0 > 0$ portion of $i S_{1 2} (P_{\bf k})$. 
However this is not the case. Within the resultant reaction-rate 
formula, which is an amplitude in RTF, the above factor 
$2 \pi n_F (p_{\bf k})\, \delta (p_{\bf k}^2) 
{P\kern-0.1em\raise0.3ex\llap{/}\kern0.15em\relax}_{\bf k}$ 
necessarily appears in association with a thermal fermion loop (see 
below for detail). The thermal fermion loop carries an extra minus 
sign, so that we have, for the portion under consideration, 
\[ 
i S_{1 2}^{(+)} (P_{\bf k}) = 2 \pi [- n_F (p_{\bf k})] \delta 
(P_{\bf k}^2) \, 
{P\kern-0.1em\raise0.3ex\llap{/}\kern0.15em\relax}_{\bf k} \, . 
\] 

Adding the contribution from the antifermion mode with 
$\{ \overline{i}_{- {\bf k}}^{(\sigma)} = \overline{j}_{- {\bf 
k}}^{(\sigma)} = 0, \; \overline{i}_{- {\bf k}}^{(\sigma) '} = 
\overline{j}_{- {\bf k}}^{(\sigma) '} = 1 \}$ ($\sigma = 
\pm$), we extract 
\begin{equation} 
2 \pi [\theta (- p_0) - n_F (p_{\bf k}) ] \, 
{P\kern-0.1em\raise0.3ex\llap{/}\kern0.15em\relax}_{\bf k} \, \delta 
(P_{\bf k}^2) \equiv i S_{1 2} (P_{\bf k}) \, . 
\end{equation} 

In the process of deduction, $i S_{j l}$ $(j, \, l = 1, 2)$ appears 
in succession. At the final stage, sets of $\langle W \rangle = 
\langle S^* S \rangle$ turn out to be thermal amplitudes $A$'s (cf. 
Eq (\ref{rate})), which includes thermal loops of the fermion 
$\psi$. Out of $A$'s, we take a \lq\lq standard'' $A_s$: Each 
fermion loop contains at most one $i S_{1 2}$. (Note that the number 
of $i S_{2 1}$ in a fermion loop is equal to the number of 
$i S_{1 2}$.) From $A_s$, we take two fermion loops $L_1$ and $L_2$ 
and let $i S_{2 1} (P) \in L_1$ and $i S_{2 1} (Q) \in L_2$. 
$i S_{2 1} (P) \, i S_{2 1} (Q)$ comes, with obvious notation, from 
$S^* S = S^* ({\bf p}, \, {\bf q}, ...) \, S ({\bf p}, \, {\bf q}, 
...) \equiv W_s$, where $S$ is the $S$-matrix element obtained using 
Feynman rules (in vacuum theory). The $S$-matrix element which is 
related to $S ({\bf p}, \, {\bf q}, ...)$ through exchange ${\bf p} 
\leftrightarrow {\bf q}$ is $- S ({\bf q}, \, {\bf p}, ...)$, where 
$S ({\bf q}, \, {\bf p}, ...)$ is obtained using Feynman rules. 
Then, we have 
\begin{equation} 
W_s \to W = - S^* ({\bf p}, \, {\bf q}, ...) \, S ({\bf q}, \, 
{\bf p}, ...) \, , 
\label{ex} 
\end{equation} 
which brings in an extra minus sign into the corresponding thermal 
amplitude $A$. Observe here that, through the above replacement of 
$S$, $L_1$ and $L_2$ in $A_s$ turns out to be an one thermal fermion 
loop $L$ in $A$. A thermal fermion loop carries a minus sign. Then 
$L_1$ and $L_2$ in $A_s$ carries $+ = (-)^2$ while $L$ in $A$ 
carries $-$. In reducing $\langle W \rangle$ to $A$, the extra minus 
sign in Eq.~(\ref{ex}) eliminates one $-$, being present in $A_s$, 
and is left with one $-$, which is interpreted as the minus sign 
associated with $L$ in $A$, What we have shown is that $A$ is a 
\lq\lq right thermal amplitude.'' 

Repeating the above procedure for \lq\lq parent'' $A_s$'s and 
\lq\lq children'' $A$'s, as \lq\lq constructed'' above, we can 
exhaust all $A$'s that contributes to the reaction-rate formula, 
and see that they are \lq\lq right'' thermal amplitudes. 
\subsection{Analysis of mode-overlapping diagrams} 
Let us turn to analyze the mode-overlapping diagrams. Noting that 
$n_{\bf k}^{(\sigma)}$ etc. and then also $i_{\bf k}^{(\sigma)}$ 
etc. take two values $0$ and $1$, we shall exhaust all the 
mode-overlapping configurations. 

(a) $\{ i_{\bf k}^{(\sigma)} = i_{\bf k}^{(\sigma) '} = j_{\bf 
k}^{(\sigma)} = j_{\bf k}^{(\sigma) '} = 1 \}$ $(\sigma = 
\pm)$ and its relatives. 

From Eq.~(\ref{fern}), $N_f = (n^{(\sigma)})^4$ ($\sigma = \pm$), 
which leads to $\langle (n^{(\sigma)})^4 \rangle = n_F$. Through by 
now familiar manner, we extract 
\begin{eqnarray} 
& & n_F \sum_{\sigma = \pm} \left[ \left\{ 2 \pi \theta (p_0) \, 
u_{j}^{(\sigma)} (P) \, \overline{u}_{j'}^{(\sigma)} (P) \right\} 
\right. \nonumber \\ 
& & \left. \mbox{\hspace*{5ex}} \times \left\{ 2 \pi \theta 
(p_0) \, u_{i}^{(\sigma)} (P) \, \overline{u}_{i'}^{(\sigma)} (P) 
\right\} \right] \, . 
\label{ya} 
\end{eqnarray} 
$u_{i}^{(\sigma)}$ and $\overline{u}_{i'}^{(\sigma)}$ 
[$u_{j}^{(\sigma)}$ and $\overline{u}_{j'}^{(\sigma)}$] in 
Eq.~(\ref{ya}) are attached to the vertices in $S$ [$S^*$]. 

The relatives, to be analyzed, of the above configuration are 
$\{ i_{\bf k}^{(\sigma)} = i_{\bf k}^{(\sigma) '} = 
j_{\bf k}^{(- \sigma)} = j_{\bf k}^{(- \sigma) '} = 1 \}$ and 
$\{ i_{\bf k}^{(\sigma)} = i_{\bf k}^{(- \sigma) '} = 
j_{\bf k}^{(\sigma)} = j_{\bf k}^{(- \sigma) '} = 1 \}$ 
($\sigma = \pm$). The former yields 
\begin{eqnarray} 
& & n_F^2 \sum_{\sigma = \pm} \left[ \left\{ 2 \pi \theta (p_0) \, 
u_{j}^{(- \sigma)} (P) \, 
\overline{u}_{j'}^{(- \sigma)} (P) \right\} \right. \nonumber \\ 
& & \left. \mbox{\hspace*{5ex}} \times \left\{ 2 \pi \theta (p_0) \, 
u_{i}^{(\sigma)} (P) \, \overline{u}_{i'}^{(\sigma)} (P) \right\} 
\right] \, , 
\label{ya1} 
\end{eqnarray} 
and the latter yields 
\begin{eqnarray} 
& & n_F (1 - n_F) \sum_{\sigma = \pm} \left[ \left\{ 2 \pi \theta 
(p_0) \, u_{j}^{(- \sigma)} (P) 
\, \overline{u}_{j'}^{(\sigma)} (P) \right\} \right. \nonumber \\ 
& & \left. \mbox{\hspace*{5ex}} \times \left\{ 2 \pi \theta (p_0) \, 
u_{i}^{(\sigma)} (P) \, \overline{u}_{i'}^{(- \sigma)} (P) \right\} 
\right] \, . 
\label{ya2} 
\end{eqnarray} 
Adding Eqs.~(\ref{ya}) and (\ref{ya1}), we obtain 
\begin{eqnarray} 
& & \left( i S_{2 2}^{(T)(+)} (P) \right)_{j j'} \left( 
i S_{1 1}^{(T)(+)} (P) \right)_{i i'} \nonumber \\ 
& & \mbox{\hspace*{5ex}} + n_F (1 - n_F) \sum_{\sigma = \pm} 
\left[ \left\{ 2 \pi 
\theta (p_0) \, u_{j}^{(\sigma)} (P) \, \overline{u}_{j'}^{(\sigma)} 
(P) \right\} \right. \nonumber \\ 
& & \left. \mbox{\hspace*{5ex}} \times \left\{ 2 \pi \theta (p_0) \, 
u_{i}^{(\sigma)} (P) \, \overline{u}_{i'}^{(\sigma)} (P) \right\} 
\right] \, . 
\label{ya3} 
\end{eqnarray} 
Adding Eqs.~(\ref{ya2}) and (\ref{ya3}), we have 
\begin{eqnarray} 
& & \left( i S_{2 2}^{(T)(+)} (P_{\bf k}) \right)_{j j'} 
\left( i S_{1 1}^{(T)(+)} (P_{\bf k}) \right)_{i i'} \nonumber \\ 
& & \mbox{\hspace*{5ex}} - \left( i S_{1 2}^{(+)} (P_{\bf k}) 
\right)_{i j'} 
\left( i S_{2 1}^{(+)} (P_{\bf k}) \right)_{j' i} \, . 
\label{ya4} 
\end{eqnarray} 
Recalling the fact that $i$ and $i'$ [$j$ and $j'$] attach to the 
vertices in $S$ [$S^*$], we see that Eq.~(\ref{ya4}) is just a 
portion of \lq\lq right'' thermal amplitude in RTF. Adding an 
appropriate sets of relative diagrams, we can extract 
Eq.~(\ref{ya4}) with complete $S$'s. 

(b) $\{ i_{\bf k}^{(+)} = i_{\bf k}^{(-)} = j_{\bf k}^{(+)} = 
j_{\bf k}^{(-)} = 1 \}$ and its relatives. 

Taking care of the anticommutativity of fermion fields, we extract 
\begin{eqnarray} 
& & n_F^2 \left[ 2 \pi \theta (p_0) \left\{  
u_{i_1}^{(+)} (P) \, u_{i_2}^{(-)} (P) - 
u_{i_2}^{(+)} (P) \, u_{i_1}^{(-)} (P) 
\right\} \right] \nonumber \\ 
& & \mbox{\hspace*{4ex}} \times \left[ 2 \pi \theta (p_0) \left\{ 
\overline{u}_{j_1}^{(+)} (P) \, \overline{u}_{j_2}^{(-)} (P) - 
\overline{u}_{j_2}^{(+)} (P) \, \overline{u}_{j_1}^{(-)} (P) 
\right\} \right] \, . 
\label{alp} 
\end{eqnarray} 
Here $u_i$'s [$\overline{u}_j$'s] are attached to the vertices in 
$S$ [$S^*$]. Simple manipulation yields 
\begin{eqnarray} 
\mbox{Eq.~(\ref{alp})} & = & \left( i S_{1 2}^{(+)} (P_{\bf k}) 
\right)_{i_1 j_1} \left( i S_{1 2}^{(+)} (P_{\bf k}) 
\right)_{i_2 j_2} \nonumber \\ 
& & - \left( i S_{1 2}^{(+)} (P_{\bf k}) \right)_{i_1 j_2} 
\left( i S_{1 2}^{(+)} (P_{\bf k}) \right)_{i_2 j_1} \, . 
\label{bet} 
\end{eqnarray} 
Adding appropriate relative diagrams, we can extract Eq.~(\ref{bet}) 
with complete $S$'s, which sits on the \lq\lq right seat'' in 
thermal amplitude in RTF (cf. Eq.~(\ref{rate})). 

(c) $\{ i_{\bf k}^{(+)} = i_{\bf k}^{(-)} = j_{\bf k}^{(+)} = 
j_{\bf k}^{(-)} = 1, \; i_{\bf k}^{(\sigma) '} = 
j_{\bf k}^{(\sigma) '} = 1 \}$ ($\sigma =\pm$) and its 
relatives. 

We extract 
\begin{eqnarray} 
& & n_F^2 \left [2 \pi \theta (p_0) \left\{ u_{i_1}^{(+)} (P) \, 
u_{i_2}^{(-)} (P) - u_{i_2}^{(+)} (P) \, u_{i_1}^{(-)} (P) \right\} 
\right] \nonumber \\ 
& & \mbox{\hspace*{4ex}} \times \left[ 2 \pi \theta (p_0) \left\{ 
\overline{u}_{j_1}^{(+)} (P) \, \overline{u}_{j_2}^{(-)} (P) - 
\overline{u}_{j_2}^{(+)} (P) \, \overline{u}_{j_1}^{(-)} (P) 
\right\} \right] \nonumber \\ 
& & \mbox{\hspace*{4ex}} \times \sum_{\sigma = \pm} \left[ 2 \pi 
\theta (p_0) \, \overline{u}_{i_3}^{(\sigma)} (P) \, 
u_{j_3}^{(\sigma)} (P) \right] \, , 
\label{hosi} 
\end{eqnarray} 
where the spinors with suffices $i_1$, $i_2$, and $i_3$ [$j_1$, 
$j_2$, and $j_3$] are attached to the vertices in $S$ [$S^*$]. 

We shall show that 
\begin{equation} 
\mbox{Eq.~(\ref{hosi})} = {\cal S}_{j_1 j_2 j_3}^{i_1 i_2 i_3} 
(P) - {\cal S}_{j_1 j_2 j_3}^{i_2 i_1 i_3} (P)
\label{A09} 
\end{equation} 
where 
\begin{eqnarray} 
{\cal S}_{j_1 j_2 j_3}^{i_1 i_2 i_3} (P) & \equiv & 
\left( i S_{1 2}^{(+)} (P) \right)_{i_1 j_1} 
\left( i S_{1 2}^{(+)} (P) \right)_{i_2 j_2} 
\left( i S_{2 1}^{(+)} (P) \right)_{j_3 i_3} \nonumber \\ 
& & + 
\left( i S_{1 1}^{(T) (+)} (P) \right)_{i_1 i_3} 
\left( i S_{1 2}^{(+)} (P) \right)_{i_2 j_2} 
\left( i S_{2 2}^{(T) (+)} (P) \right)_{j_3 j_1} \nonumber \\ 
& & + 
\left( i S_{1 2}^{(+)} (P) \right)_{i_1 j_1} 
\left( i S_{1 1}^{(T) (+)} (P) \right)_{i_2 i_3} 
\left( i S_{2 2}^{(T) (+)} (P) \right)_{j_3 j_2} \, . 
\label{B} 
\end{eqnarray} 

\noindent We shall prove this by running in the opposite direction, 
i.e., starting from Eq.~(\ref{A09}), we derive Eq.~(\ref{hosi}). The 
first term on the RHS of Eq.~(\ref{B}) consists of two terms, the 
one is proportional to $n_F^2$ and the one is proportional to 
$n_F^3$. The second and third terms are proportional to $n_F^3$. 
$\left( i S_{1 2}^{(+)} (P) \right)_{i_1 j_1}$ may be written as 
(cf. Eq.~(\ref{D})) 
\begin{equation} 
\left( i S_{1 2}^{(+)} (P) \right)_{i_1 j_1} = - 2 \pi n_F (p) \, 
\delta (P^2) \sum_{\sigma = \pm} u_{i_1}^{(\sigma)} (P) 
\overline{u}_{j_1}^{(\sigma)} (P) \, . 
\label{dec} 
\end{equation} 
Other $S$'s in Eq.~(\ref{B}) may be expressed similarly. 
Straightforward but tedious manipulation shows that the \lq\lq 
$n_F^3$ part'' of ${\cal S}_{j_1 j_2 j_3}^{i_1 i_2 i_3} - 
{\cal S}_{j_1 j_2 j_3}^{i_2 i_1 i_3}$ vanishes. Then, in 
Eq.~(\ref{A09}), we are left with \lq\lq $n_F^2$ part'', which turns 
out to be Eq.~(\ref{hosi}). 

The same comment as above after eq. (\ref{bet}) applies here. 

(d) $\{ i_{\bf k}^{(+)} = i_{\bf k}^{(-)} = j_{\bf k}^{(+)} = 
j_{\bf k}^{(-)} = i_{\bf k}^{(+) '} = i_{\bf k}^{(-) '} = 
j_{\bf k}^{(+) '} = j_{\bf k}^{(-) '} = 1 \}$ and its 
relatives. 

We extract 
\begin{eqnarray} 
& & n_F^2 \left[ 2 \pi \theta (p_0) \left\{ 
u_{i_1}^{(+)} (P) \, u_{i_2}^{(-)} (P) - 
u_{i_1}^{(-)} (P) \, u_{i_2}^{(+)} (P) 
\right\} \right] \nonumber \\ 
& & \mbox{\hspace*{4ex}} \times 
\left[ 2 \pi \theta (p_0) \left\{ 
\overline{u}_{j_1}^{(+)} (P) \, \overline{u}_{j_2}^{(-)} (P) - 
\overline{u}_{j_1}^{(-)} (P) \, \overline{u}_{j_2}^{(+)} (P) 
\right\} \right] \nonumber \\ 
& & \mbox{\hspace*{4ex}} \times 
\left[ 2 \pi \theta (p_0) \left\{ 
\overline{u}_{i_3}^{(+)} (P) \, \overline{u}_{i_4}^{(-)} (P) - 
\overline{u}_{i_3}^{(-)} (P) \, \overline{u}_{i_4}^{(+)} (P) 
\right\} \right] \nonumber \\ 
& & \mbox{\hspace*{4ex}} \times 
\left[ 2 \pi \theta (p_0) \left\{ 
u_{j_3}^{(+)} (P) \, u_{j_4}^{(-)} (P) - 
u_{j_3}^{(-)} (P) \, u_{j_4}^{(+)} (P) \right\} 
\right] \, . 
\label{nan} 
\end{eqnarray} 
As in the above case (c), through straightforward but tedious 
calculation, we obtain 
\begin{eqnarray} 
\mbox{Eq.~(\ref{nan})} & = & 
{\cal S}_{j_1 j_2 j_3 j_4}^{i_1 i_2 i_3 i_4} (P) - 
{\cal S}_{j_1 j_2 j_3 j_4}^{i_2 i_1 i_3 i_4} (P) \nonumber \\ 
& & - {\cal S}_{j_1 j_2 j_4 j_3}^{i_1 i_2 i_3 i_4} (P) + 
{\cal S}_{j_1 j_2 j_4 j_3}^{i_2 i_1 i_3 i_4} (P) \, , 
\end{eqnarray} 
where 
\begin{eqnarray*} 
{\cal S}_{j_1 j_2 j_3 j_4}^{i_1 i_2 i_3 i_4} (P) & \equiv & 
\left( i S_{1 2}^{(+)} (P) \right)_{i_1 j_1} 
\left( i S_{1 2}^{(+)} (P) \right)_{i_2 j_2} 
\left( i S_{2 1}^{(+)} (P) \right)_{j_3 i_3} 
\left( i S_{2 1}^{(+)} (P) \right)_{j_4 i_4} \nonumber \\ 
& & + \left( i S_{1 1}^{(T) (+)} (P) \right)_{i_1 i_3} 
\left( i S_{1 2}^{(+)} (P) \right)_{i_2 j_2} 
\left( i S_{2 2}^{(T) (+)} (P) \right)_{j_3 j_1} 
\left( i S_{2 1}^{(+)} (P) \right)_{j_4 i_4} \nonumber \\ 
& & 
+ \left( i S_{1 2}^{(+)} (P) \right)_{i_1 j_1} 
\left( i S_{1 1}^{(T) (+)} (P) \right)_{i_2 i_3} 
\left( i S_{2 2}^{(T) (+)} (P) \right)_{j_3 j_2} 
\left( i S_{2 1}^{(+)} (P) \right)_{j_4 i_4} \nonumber \\ 
& & 
+ \left( i S_{1 2}^{(+)} (P) \right)_{i_1 j_1} 
\left( i S_{2 1}^{(+)} (P) \right)_{j_3 i_3} 
\left( i S_{1 1}^{(T) (+)} (P) \right)_{i_2 i_4} 
\left( i S_{2 2}^{(T) (+)} (P) \right)_{j_4 j_2} \nonumber \\ 
& & 
+ \left( i S_{1 1}^{(T) (+)} (P) \right)_{i_1 i_4} 
\left( i S_{1 2}^{(+)} (P) \right)_{i_2 j_2} 
\left( i S_{2 1}^{(+)} (P) \right)_{j_3 i_3} 
\left( i S_{2 2}^{(T) (+)} (P) \right)_{j_4 j_1} \nonumber \\ 
& & 
+ \left( i S_{1 1}^{(T) (+)} (P) \right)_{i_1 i_4} 
\left( i S_{1 1}^{(T) (+)} (P) \right)_{i_2 i_3} \nonumber \\ 
& & \times \left( i S_{2 2}^{(T) (+)} (P) \right)_{j_4 j_1} 
\left( i S_{2 2}^{(T) (+)} (P) \right)_{j_3 j_2} 
\, . 
\end{eqnarray*} 

\noindent The same comment as above after Eq.~(\ref{bet}) applies 
here. 

There remains following two configurations to be analyzed; (e) 
$\{ i_{\bf k}^{(+) '} = i_{\bf k}^{(-) '} = j_{\bf k}^{(+) '} = 
j_{\bf k}^{(-) '} = 1 \}$ and its relatives and (f) $\{ 
i_{\bf k}^{(+) '} = i_{\bf k}^{(-) '} = j_{\bf k}^{(+) '} = 
j_{\bf k}^{(-) '} = 1, \; i_{\bf k}^{(\sigma)} = 
j_{\bf k}^{(\sigma)}= 1 \}$ ($\sigma = \pm$) and its 
relatives. The case (e) [(f)] may be analyzed in a similar manner as 
(b) [(c)] above and the \lq\lq right combination'' of thermal 
propagators is extracted. 

As in the scalar-field case, Sec.~III~B, there appear disconnected 
${\cal N}$'s; ${\cal N} = {\cal N}_{\scriptsize{con}} {\cal D}$ with 
${\cal D} \neq 1$. Such cases are treated in a same manner as in the 
scalar-field case. 

This completes the analysis of all mode-overlapping configurations. 

Conversely, we take a diagram for $A$ in the reaction-rate formula 
(cf. Eq.~(\ref{rate})). The amplitude $A$ contains \lq\lq vanishing 
contributions,'' which {\em should} vanish. By this we mean the 
contributions coming from the configurations, in which at least one 
of $i_{\bf k}^{(h)}$, $i_{\bf k}^{(h) '}$, $j_{\bf k}^{(h)}$, 
$j_{\bf k}^{(h)'}$, $\overline{i}_{\bf k}^{(h)}$, 
$\overline{i}_{\bf k}^{(h) '}$, $\overline{j}_{\bf k}^{(h)}$, 
$\overline{j}_{\bf k}^{(h) '}$ ($h = \pm$) is equal to or greater 
than 2. Let us show that such contributions really vanish. Suppose 
that $A$ contains 
\begin{equation} 
\prod_{k = 1}^3 \left( i S_{1 2} (R_k) \right)_{i_k j_k} \, , 
\label{maru0} 
\end{equation} 
where $R_k$ ($k = 1, 2, 3$) is the loop momentum (cf. 
Eq.~(\ref{loop})) and the suffix \lq $i_k j_k$' stands for the 
$(i_k, j_k)$ element of $i S_{1 2}$ in the $4 \times 4$ Dirac-matrix 
space. In the loop-momentum space, there are \lq\lq points,'' where 
$R_1 = R_2 = R_3 \equiv R = (r_0, {\bf r})$. Adding the 
contributions from the five relative diagrams, we have, in place of 
Eq.~(\ref{maru0}), 
\begin{equation} 
\sum_{perm} 
\sigma^{j_1 j_2 j_3}_{l_1 l_2 l_3} 
\prod_{k = 1}^3 \left( i S_{1 2} (R) \right)_{i_k l_k} \, , 
\label{maru1} 
\end{equation} 
where summation is taken over all permutations of $(j_1 j_2 j_3)$. 
$\sigma^{j_1 j_2 j_3}_{l_1 l_2 l_3} = + / -$ when $(l_1 l_2 l_3)$ is 
an even/odd permutation of $(j_1 j_2 j_3)$, which is a reflection of 
the anticommutativity of fermion fields. We take the case $r_0 > 0$. 
The \lq\lq type-1 side'' of Eq.~(\ref{maru1}) comes from 
$i_{\bf k}^{(+)} + i_{\bf k}^{(-)} = 3$, and then $i_{\bf k}^{(+)} 
\geq 2$ or $i_{\bf k}^{(-)} \geq 2$. Then the contribution under 
consideration {\em should} vanish. In order to see that this is 
really the case, using the expression (\ref{dec}), we further 
extract from Eq.~(\ref{maru1}) 
\begin{equation} 
\sum_{perm} 
\sigma^{j_1 j_2 j_3}_{l_1 l_2 l_3} \prod_{k = 1}^3 \left[ 
\sum_{\sigma_k = \pm} u_{i_k}^{(\sigma_k)} (R) \, 
\overline{u}_{l_k}^{(\sigma_k)} (R) \right] \, . 
\label{maru3} 
\end{equation} 
Again straightforward but tedious manipulation shows that 
Eq~. (\ref{maru3}) is in fact vanishes. In a similar manner, we can 
show that Eq.~(\ref{maru1}) with $r_0 < 0$ also vanishes. 

We can also see that $\left( i S_{1 1}^{(T)} (R) 
\right)_{i_1 j_1} \prod_{k = 2}^3 \left( i S_{1 2} (R) 
\right)_{i_k j_k}$ and its relatives add up to vanish. When product 
of $n \, (\geq 4)$ $i S_{1 2} (R)$ and/or $i S_{1 1}^{(T)} (R)$ 
appears in $A$, pick out three of them and apply the above argument 
to show that the contribution vanishes. 

The above analysis applies to all other \lq\lq vanishing 
contributions,'' which include $\prod_{k = 1}^3 \left( i S_{2 1} 
(R_k) \right)$ with its relatives etc. This completes the proof of 
absence of \lq\lq vanishing contributions.'' 
\setcounter{equation}{0}
\setcounter{section}{4}
\section{The rate of reactions between the constituent particles 
of the heat bath} 
\def\theequation{\mbox{\arabic{section}.\arabic{equation}}}
In the heat bath composed of scalar fields $\phi$'s, taking place 
is the reaction, 
\begin{eqnarray} 
& & \phi ({\bf p}_1) + ... + \phi ({\bf p}_m) + \mbox{heat bath} 
\nonumber \\ 
& & \mbox{\hspace*{4ex}} \to 
\phi ({\bf q}_1) + ... + \phi ({\bf q}_n) + \mbox{anything} \, , 
\label{the-re} 
\end{eqnarray} 
where $\phi$'s are the constituent particle of the heat bath. 
One can easily show that the reaction rate takes the form, 
\begin{eqnarray}
\frac{1}{V} \left( \prod_{j = 1}^n 2 q_j V \right) 
{\cal R} & = & \left( \prod_{i = 1}^m \frac{1}{2 
p_i V} \right) \left( \prod_{i = 1}^m n_B (p_i) 
\right) \left( \prod_{j = 1}^n \{ 1 + n_B (q_j) \} \right) 
\nonumber \\ 
& & \times A (P_1^{(2)}, ..., P_m^{(2)}, 
Q_1^{(1)}, ..., Q_n^{(1)}; P_1^{(1)}, ..., P_m^{(1)}, Q_1^{(2)}, 
..., Q_n^{(2)}) \, , \nonumber \\ 
\label{rate1} 
\end{eqnarray}

\noindent where $A$ is the RTF amplitude for the forward process, 
\begin{eqnarray}
& & \phi_1 (P_1) + ... + \phi_1 (P_m) + \phi_2 (Q_1) + ... + 
\phi_2 (Q_n) \nonumber \\ 
& & \mbox{\hspace*{5ex}} \to \phi_2 (P_1) + ... + \phi_2 (P_m) + 
\phi_1 (Q_1) + ... + \phi_1 (Q_n) \, . 
\label{reaction1} 
\end{eqnarray}

It is worth noting that Eq.~(\ref{rate1}) may be rewritten as 
\begin{eqnarray}
\frac{1}{V} \, {\cal R} & = & 
\left[ \prod_{i = 1}^m \frac{1}{V} 
\int \frac{d p_{i 0}}{2 \pi} \, \theta (p_{i 0}) \, 
i D_{1 2} (P_i) \right] \nonumber \\ 
& & \times 
\left[ \prod_{j = 1}^n \frac{1}{V} 
\int \frac{d q_{j 0}}{2 \pi} \, \theta (q_{j 0}) \, 
i D_{2 1} (Q_j) \right] \, A  \nonumber \\ 
& \equiv & \tilde{A}_{\scriptsize{bubble}} \, . 
\end{eqnarray}
The RHS, $\tilde{A}_{\scriptsize{bubble}}$, is a no-leg thermal 
amplitude, in which no summation is taken over ${\bf p}_i$ ($i = 1, 
..., m$) and ${\bf q}_j$ ($j = 1, ..., n$). 

Generalization of the above result to the theories with gauge bosons 
and/or fermions is straightforward. 
\setcounter{equation}{0}
\setcounter{section}{5}
\section{Detailed balance} 
\def\theequation{\mbox{\arabic{section}.\arabic{equation}}}
In this section, on the basis of the generalized reaction-rate 
formula, Eq.~(\ref{rate1}), we derive the detailed-balance formula 
{\em through diagrammatic analysis}. 

The purpose of this section is to show that the rate (\ref{rate1}) 
for the process (\ref{the-re}) is equal to the rate for the inverse 
process to (\ref{the-re}). [For the case of theories with gauge 
bosons and/or fermions, the same result 
is obtained.] This is well known for the cases of decay- and 
production-processes, which corresponds to $m = 1$, $n = 0$ and 
$m = 0$, $n = 1$, respectively, in Eq.~(\ref{rate1}). 

Take a diagram for $A$, Eq.~(\ref{rate1}), and let $N_1$ and $N_2$ 
be the number of $i D_{2 1}$'s and $i D_{1 2}$'s, respectively, 
which is involved in $A$, 
\begin{equation} 
\prod_{j = 1}^{N_1} i D_{2 1} (R_j) \, \prod_{k = 1}^{N_2} i D_{1 2} 
(R_{N_1 + k}) \, . 
\label{12} 
\end{equation} 
By cutting all the lines $i D_{1 2}$'s and $i D_{2 1}$', we divide 
$A$ into one or several \lq\lq type-1 islands'' and one or several 
\lq\lq type-2 islands''. 
Here, the type-1 (type-2) 
island is a \lq\lq maximal'' {\em amputated} subdiagram of $A$, 
which consists of only type-1 (type-2) vertices and of the 
propagators $i D_{1 1}$'s ($i D_{2 2}$'s) connecting them. Then, a 
type-1 (type-2) island includes no type-2 (type-1) vertex. A type-1 
(type-2) island is connected by $i D_{2 1}$'s and/or $i D_{1 2}$'s 
to type-2 (type-1) island(s). 

Take a type-1 island and we write its contribution (to $A$) 
\begin{equation} 
{\cal I}_1 (Q_{j_1}, \, ..., \, Q_{j_{\ell'}}; \, P_{i_1}, \, ..., 
\, P_{i_\ell}) \, . 
\label{i1} 
\end{equation} 
Here $\{ P_{i_k}, \, 1 \leq k \leq \ell \}$ is a subset of $\{ 
P_{i}, \, 1 \leq i \leq m \}$ on the LHS of Eq.~(\ref{reaction1}) 
and $\{ Q_{j_k}, \, 1 \leq k \leq \ell' \}$ is a subset of $\{ Q_j, 
\, 1 \leq j \leq n \}$ on the RHS of Eq.~(\ref{reaction1}), where 
$\ell$, $\ell'$ $\geq 0$. This type-1 island is connected by $s_1 
(\geq 0)$ propagators $i D_{2 1}$'s and $s_2 (\geq 0)$ propagators 
$i D_{1 2}$'s to one or several type-2 islands. With the help of the 
identity, 
\begin{equation} 
D_{2 1} (R) = e^{\beta r_0} \, D_{1 2} (R) \, , 
\label{yoku} 
\end{equation} 
and the momentum-conservation condition, we obtain, for $i D$'s that 
are attached to ${\cal I}_1$, 
\begin{eqnarray} 
& & \prod_{j = 1}^{s_1} i D_{2 1} (R_j) \prod_{k = 1}^{s_2} 
i D_{1 2} (R_{s_1 + k}) \nonumber \\ 
& & \mbox{\hspace*{4ex}} = \mbox{exp} \left( \beta \left[ 
\sum_{k = 1}^\ell 
p_{i_k} - \sum_{k = 1}^{\ell'} q_{j_k} \right]  
\right) \prod_{j = 1}^{s_1} i D_{1 2} (R_j) \prod_{k = 1}^{s_2} i 
D_{2 1} (R_{s_1 + k}) \, . 
\label{il1} 
\end{eqnarray} 

We now take a type-2 island, whose contribution is written as 
\begin{equation} 
{\cal I}_2 (P_{i_1}, \, ..., \, P_{i_{\ell}}; \, Q_{j_1}, \, ..., \, 
Q_{j_{\ell'}}) \, , 
\label{i2} 
\end{equation} 
where $\{ Q_{j_k}, \, 1 \leq k \leq \ell' \}$ is a subset of $\{ 
Q_j, \, 1 \leq j \leq n \}$ on the LHS of Eq.~(\ref{reaction1}) 
and $\{ P_{i_k}, \, 1 \leq k \leq \ell \}$ is a subset of $\{ P_i, 
\, 1 \leq i \leq m \}$ on the RHS of Eq.~(\ref{reaction1}). $\ell$ 
($\ell'$) here is not necessarily equal to $\ell$ ($\ell'$) in 
Eq.~(\ref{i1}). In a similar manner as above, in place of 
Eq.~(\ref{il1}), we have, with obvious notation, 
\begin{eqnarray} 
\prod_{j = 1}^{s_1'} i D_{2 1} (R_j) \prod_{k = 1}^{s_2'} i D_{1 2} 
(R_{s_1' + k}) & = & \mbox{exp} \left( \beta \left[ 
\sum_{k = 1}^\ell p_{i_k} - \sum_{k = 1}^{\ell'} q_{j_k} \right]  
\right) \nonumber \\ 
& & \times \prod_{j = 1}^{s_1'} i D_{1 2} (R_j) \prod_{k = 1}^{s_2'} 
i D_{2 1} (R_{s_1' + k}) \, . 
\label{il2} 
\end{eqnarray} 

For all the islands, we make the above replacements, i.e., the LHS 
of Eqs.~(\ref{il1}) and (\ref{il2}) are replaced with respective 
RHS. Through this procedure, each $i D_{2 1}$ and each $i D_{1 2}$ 
in Eq.~(\ref{12}) is \lq\lq used'' twice. Then we obtain  
\begin{eqnarray*} 
\mbox{Eq.~(\ref{12})} & = & \mbox{exp} \left( \beta \left[ 
\sum_{j = 1}^{m} p_j - \sum_{j = 1}^{n} q_j \right] \right) \\ 
& & \times \prod_{j = 1}^{N_1} i D_{1 2} (R_j) \prod_{k =1}^{N_2} i 
D_{2 1} (R_{N_1 + k}) \, . 
\end{eqnarray*} 

Now we note that the propagators in ${\cal I}_1$'s (${\cal I}_2$'s) 
are $i D_{1 1}$'s ($i D_{2 2}$'s), and vertices in ${\cal I}_1$'s 
(${\cal I}_2$'s) are $i \lambda$ ($- i \lambda$) [cf. above after 
Eq.~(\ref{i-l})]. Then, using the relation (\ref{d22}), 
\[ 
\left[ i D_{1 1} (R) \right]^* = i D_{2 2} (R) \, , 
\] 
and $\left[ i \lambda \right]^*$ $=$ $- i \lambda$, we easily see 
that\footnote{A comment on QCD (QED) is in order. As to the 4-gluon 
vertex, when compared to the scalar theory, no new feature arises. 
Let ${\cal V}_i$ ($i = 1, 2$) be the factor that is associated to a 
trigluon vertex in a type-$i$ island. ${\cal V}_i$ is real and 
${\cal V}_2 = - {\cal V}_1$. Then, in place of 
Eq.~(\ref{saigo-mae}), we have ${\cal I}_1{}^* = (-)^{N} {\cal I}_2$ 
with $N$ the number of trigluon vertices in ${\cal I}_1$. Since $A$ 
in Eq.~(\ref{rate1}) contains even number of trigluon vertices, 
Eq.~(\ref{saigo}) holds unchanged. Let us turn to analyze the 
quark-gluon vertex. In a standard notation, the factor associated to 
a quark-gluon vertex in a type-1/2 island is $\pm i g \gamma^\mu 
T^a$. Taking trace, in $A$ in Eq.~(\ref{rate1}), of the products of 
$\gamma$-matrices and of color matrices yield a real function of 
$P$'s and $Q$'s. Then, $(i g)^* = - i g$ leads to Eq.~(\ref{saigo}). 
To sum up, Eq.~(\ref{saigo}) holds for QCD (QEC).} 
\begin{eqnarray} 
& & \left[ {\cal I}_1 (Q_{j_1}, \, ..., \, Q_{j_{\ell'}}; \, 
P_{i_1}, \, ..., \, P_{i_\ell}) \right]^* \nonumber \\ 
& & \mbox{\hspace*{4ex}} = {\cal I}_2 (Q_{j_1}, \, ..., \, 
Q_{j_{\ell'}}; \, P_{i_1}, \, ..., \, P_{i_\ell}) \, .  
\label{saigo-mae} 
\end{eqnarray} 

Here we note that, from the first-principle derivation above, it is 
obvious that, to any order of perturbation series, the amplitude $A$ 
in Eq.~(\ref{rate1}) is real, provided that all the contributing 
diagrams are added. This fact, together with Eq.~(\ref{saigo-mae}), 
shows that 
\begin{eqnarray} 
& & A (P_1^{(2)}, ..., P_m^{(2)}, Q_1^{(1)}, ..., Q_n^{(1)}; 
P_1^{(1)}, ..., P_m^{(1)}, Q_1^{(2)}, ..., Q_n^{(2)}) \nonumber 
\\ 
& & \mbox{\hspace*{5ex}} = \mbox{exp} \left( \beta \left[ 
\sum_{i = 1}^{m} p_i - \sum_{j = 1}^{n} q_j \right] \right) 
\nonumber \\ 
& & \mbox{\hspace*{8ex}} \times A (Q_1^{(2)}, ..., Q_n^{(2)}, 
P_1^{(1)}, ..., 
P_m^{(1)}; Q_1^{(1)}, 
..., Q_n^{(1)}, P_1^{(2)}, ..., P_m^{(2)}) \, . 
\label{saigo} 
\end{eqnarray} 

Using Eq.~(\ref{yoku}), we obtain 
\begin{eqnarray} 
e^{\beta p_i} n_B (p_i) & = & 1 + n_B (p_i) 
\, , \nonumber \\ 
e^{- \beta q_j} \{ 1 + n_B (q_j) \} & = & n_B 
(q_j) \, . 
\label{mada} 
\end{eqnarray} 
Substituting Eq.~(\ref{saigo}) into Eq.~(\ref{rate1}) and using 
Eq.~(\ref{mada}), we finally obtain 
\begin{equation} 
\frac{1}{V} \left( \prod_{j = 1}^n 2 q_j V \right) {\cal R} 
= \frac{1}{V} \left( \prod_{i = 1}^m 2 p_i V \right) 
{\cal R'} \, . 
\label{final} 
\end{equation} 
Here, the LHS is the rate of the thermal reaction (\ref{the-re}) 
while the RHS is the rate of its inverse process 
\begin{eqnarray*} 
& & \phi ({\bf q}_1) + ... + \phi ({\bf q}_n) + \mbox{heat bath} 
\\ 
& & \mbox{\hspace*{4ex}} \to 
\phi ({\bf p}_1) + ... + \phi ({\bf p}_m) + \mbox{anything} \, . 
\end{eqnarray*} 
Equation (\ref{final}) is the desired detailed-balance formula. 
\setcounter{equation}{0}
\setcounter{section}{6}
\section{$T \to 0$ limit and Cutkosky rules} 
\def\theequation{\mbox{\arabic{section}.\arabic{equation}}}
In this section, we show that, in the limit, $T \to 0$, the 
reaction-rate formula (\ref{rate}) reduces to the formula that is 
obtained using the Cutkosky rules. Then, in the case 
of $m = 2$ and $n = 0$, Eq.~(\ref{rate}) goes to the optical theorem 
and, for $m = 2$ and $n = 1$, Eq.~(\ref{rate}) goes to the Mueller 
formula \cite{mue} for inclusive reactions.  

In the previous section, for a given diagram for $A$ in 
Eq.~(\ref{rate}), we have defined a set of \lq\lq islands''. The 
islands in the set may be classified in two groups. The first group 
consists of the islands, which contains at least one external 
vertex. Here the external vertex is the vertex, in which or from 
which the external momentum flows. The second group consists of the 
isolated islands, which have no external vertex. 

Let us take the scalar field theory and investigate 
zero-temperature limit ($T \to 0$) of the 
reaction-rate formula, Eq.~(\ref{rate}). 
[Again, generalization to other theories is straightforward.] 
In this limit, $i D_{2 1} 
(P)$ $\to$ $2 \pi \theta (p_0) \, \delta (P^2)$ and $i D_{1 2} (P)$ 
$\to$ $2 \pi \theta (- p_0) \, \delta (P^2)$. It can readily be seen 
that, due to momentum conservation, ${\cal I}_1$ and ${\cal I}_2$, 
Eqs.~(\ref{i1}) and (\ref{i2}), corresponding to the isolated 
islands vanish. Then, the nonvanishing amplitude $A$ contains only 
the islands belonging to the first group. Thus, we obtain 
\begin{eqnarray} 
A & = & \prod_{j = 1}^s \left[ 2 \pi \theta (r_{j 0}) \, \delta 
(R_j^2) \right] \prod_{i = 1}^{N_1} {\cal I}_1 (\{ P \}_i; \, \{ Q 
\}_i) \nonumber \\ 
& & \times \prod_{j = 1}^{N_2} {\cal I}_2 (\{ Q \}_j; \, 
\{ P \}_j) \, , 
\label{kut} 
\end{eqnarray} 
where $\{ P \}_i$ etc. denotes the subset of $P_1, ..., \, P_m$, 
which flow in the $i$th \lq\lq type-1 island'' etc. $\{ P \}_i \cup 
\{ Q \}_i$ and $\{ Q \}_j \cup \{ P \}_j$ are not empty. In 
Eq.~(\ref{kut}), the direction of all the $s$ momenta, $R$'s, each 
of which connects a \lq\lq type-1 island'' and a \lq\lq type-2 
island,'' is taken to flow from the \lq\lq type-1 island'' to the 
\lq\lq type-2 island''. As noted before, the diagram representing 
$A$ in Eq.~(\ref{kut}) is connected. 

The RHS of Eq.~(\ref{kut}) is just the quantity, which is obtained 
by applying the Cutkosky rules \cite{cut} (in vacuum theory) to the 
present case. As a special case, consider Eq.~(\ref{kut}) with 
$m = 2$ and $n = 0$. Since the particle represented by $\phi$ is 
stable at $T = 0$, 
in Eq.~(\ref{kut}), $N_1$ $=$ $N_2$ $= 1$ and $\{ P \}_{i = 1}$ $=$ 
$\{ P \}_{j = 1}$ $=$ $\{ P_1, \, P_2 \}$. Thus Eq.~(\ref{kut}) is 
the optical theorem in vacuum theory. Similarly, for $m = 2$ and $n 
\geq 1$, Eq.~(\ref{kut}) is just the (generalized) Mueller formula 
\cite{mue} for the inclusive process, 
\[ 
\Phi ({\bf p}_1) + 
\Phi ({\bf p}_2) \to \Phi ({\bf q}_1) + ... + \Phi ({\bf q}_n) + 
\mbox{anything} \, . 
\] 
\setcounter{equation}{0}
\setcounter{section}{7}
\section{Thermal cutting rules} 
\def\theequation{\mbox{\arabic{section}.\arabic{equation}}}
In view of controversy mentioned in Sec.~I, we survey in this 
section the discussions made in the past for the thermal Cutkosky 
formula and thermal cutting rules. Although no new result is 
involved here, it is worth pigeonholing the issue. 
The Cutkosky formula \cite{cut} in vacuum theory is the formula that 
relates the imaginary or absorptive part of an amplitude $A$ to the 
sum of cut amplitudes $\sum_{\mbox{\scriptsize{cuts}}} 
B^{\mbox{{\scriptsize{(cut)}}}}$. For simplicity, in this section, 
we take a self-interacting complex scalar field theory. 
Generalization to other theories are straightforward. 
$B^{\mbox{{\scriptsize{(cut)}}}}$'s are constructed from $A$ by so 
cutting the propagators $i D$'s in $A$ that $A$ is divided into 
$A_S$ and $A_{S^*}$, which are amputated. Here $A_S$ is a part(s) 
of $A$ and $A_{S^*}$ is the complex conjugate of the amplitude that 
is obtained from $A$ by removing $A_S$ and $i D$'s. Cutting the 
propagator $i D (P)$ makes $i D (P)$ 
\begin{equation} 
2 \pi \theta (\pm p_0) \, \delta (P^2 - m^2) \, , 
\label{cut-0} 
\end{equation} 
where the upper (lower) sign is taken when $P$ flows from a vertex 
in $A_S$ ($A_{S^*}$) to a vertex in $A_{S^*}$ ($A_S$). When the 
Cutkosky formula is applied to a forward amplitude $A$, we see that 
$Im A$ is proportional to the corresponding reaction rate, where 
cutted propagators represent the (on-shell) particles in the final 
state. 

Kobes and Semenoff (KS) \cite{kob} were the first who generalized 
the Cutkosky formula to the case of RTF. Namely they obtained the 
formula that relates the imaginary part of a thermal amplitude to 
the sum of \lq\lq circled amplitudes,'' each of which corresponds to 
the \lq\lq circled'' diagram that includes the so-called circled and 
uncircled vertices. The first paper of \cite{kob} discusses general 
thermal amplitudes and the second one discusses physical amplitudes, 
i.e., amplitudes with all external vertices being of type 1. In the 
sequel, unless otherwise stated, we shall restrict our concern to 
the physical amplitudes. The thermal Cutkosky formula deduced in 
\cite{kob} may be written in terms of thermal amplitudes in RTF: 
\begin{eqnarray} 
& & Im \left[ i G 
(P_1^{(1)}, \cdots, P_n^{(1)}) \right] \nonumber \\ 
& & \mbox{\hspace*{5ex}} = - \frac{1}{2} \sum_{i_1, \, ..., \, 
i_n = 1}^{2} \mbox{\hspace*{-3.0ex}} \raisebox{1.2ex}{$'$} \, 
G (P_1^{(i_1)}, \cdots, P_n^{(i_n)}) \, . 
\label{cut-t} 
\end{eqnarray} 
Here $G (P_1^{(i_1)}, \cdot \cdot \cdot, P_n^{(i_n)})$ stands for 
the (amputated) thermal amplitude with type-$i_j$ ($j = 1, \, ..., 
n$) external vertex in which or from which $P_j$ flows. In 
Eq.~(\ref{cut-t}), the sum $\sum'$ stands for taking summation 
excluding $i_1 = ... = i_n = 1$ and $i_1 = ... = i_n = 2$. Note 
that, as a matter of course, in $G$, sum is taken over the types (1 
and 2) for all internal vertices. 

KS then generalized the notion of cuttings. Comparison of $i D_{2 1} 
(P)$, Eq.~(\ref{21}), and $i D_{12} (P) = i D_{2 1} (- P)$, 
Eq.~(\ref{aha}), with Eq.~(\ref{cut-0}) leads them to regard $i 
D_{1 2}$ and $i D_{2 1}$ in $G$'s on the RHS of Eq.~(\ref{cut-t}) as 
{\em cutted propagators}. Through cuttings, each $G$ is divided into 
several pieces. KS then introduced a notion of {\em cuttabe} and 
{\em uncuttable diagrams}. The former diagram is the diagram that 
does not include isolated island(s) (cf. Sec.~VII) while the latter 
diagram includes at least one isolated island. Note that, in the 
case of vacuum theory, all the diagrams are cuttable ones, which 
motivates KS to introduce the above definition. Thus, the 
terminology \lq\lq uncuttable'' sounded quite natural at the time of 
its introduction. In spite of the fact that this is a matter of 
definition, existence of {\em uncuttable diagrams} has aroused 
controversy. 

Kobes analyzed \cite{kob1} retarded Green functions in terms of 
circled diagrams. As to the usage of \lq\lq cuttings'', \lq\lq 
cuttable'', and \lq\lq uncuttable,'' he followed \cite{kob}. 

Jeon analyzed \cite{jeo} two-point functions in imaginary-time 
formalism. Continuing to the real energies, he discussed thermal 
cutting rules. His definition of cutting is the same as in 
\cite{kob}, i.e., the propagators $i D_{1 2}$ and $i D_{2 1}$ are 
regarded as cutted propagators. No mention was made on the cuttable 
and uncuttable diagrams, but no doubt that he supposed all diagrams 
to be cuttable. 

Bedeque, Das, and Naik analyzed \cite{bed} the imaginary part of 
thermal amplitudes (physical and \lq\lq unphysical) from the same 
starting formula as in \cite{kob}, but with different route. Recall 
that the propagator $i D_{j k}$ ($j, \, k = 1, 2)$ connects a 
type-$j$ vertex with a type-$k$ vertex. $i D_{j k}$ is defined to be 
a {\em cutted propagator} if and only if one of the type-$j$ and 
type-$k$ vertices is of circled and another is of uncircled (cf. the 
first paper of \cite{kob}). They then showed that the imaginary part 
of a thermal amplitude is written as the sum of {\em cuttable 
diagrams}, in the sense of KS stated above. In each cuttable 
diagram, connected subdiagram(s) at one side of the cut line 
contains only uncircled vertices (external and internal) while 
connected subdiagram(s) at the other side of the cut line contains 
only circled vertices. As was pointed out in \cite{gel}, however, 
each connected part contains in general propagators that are 
proportional to the on-shell factor $\delta (P^2 - m^2)$. Of course, 
in the zero-temperature limit, their formula as well as KS's one 
reduce to the Cutkosky formula. 

Gelis extensively analyzed \cite{gel} thermal cutting rules for 
various formulations of real-time thermal field theory. As to the 
usage of \lq\lq cuttings'', \lq\lq cuttable'', and 
\lq\lq uncuttable,'' he followed \cite{kob}. 

Cutting rules for thermal reaction-rate formula are discussed in 
\cite{nie,nie-1,nie-2,nie1,nie-tak,kus}. Note that, as mentioned 
above, in vacuum theory, the cutted propagator, Eq.~(\ref{cut-0}), 
corresponds to the (on-shell) final-state particle. The thermal 
cutting rules introduced in \cite{nie,nie-1,nie-2,nie1,nie-tak,kus} 
is a generalization of this fact. As we have seen above, $i G_{1 2}$ 
(which collectively denotes $i D_{1 2}$ and $i S_{1 2}$) 
[$i G_{2 1}$] consists of two parts, the one comes from the 
particle [antiparticle] in the initial state and another comes from 
the antiparticle [particle] in the final state. While $i 
G_{1 1}^{(T)}$ and $i G_{2 2}^{(T)}$, the $T$-dependent parts of 
$i G_{1 1}$ and $i G_{2 2}$, come from the interplay of 
initial-state (anti)particle and the final-state (anti)particle. We 
recall that each of the thermal propagators $i G_{1 1}$ and 
$i G_{2 2}$ consists of two parts, the $T = 0$ part $i G^{(0)}$ and 
the $T$-dependent part $i G^{(T)}$. Then, $A$ in Eq.~(\ref{rate}) or 
(\ref{rate1}) is divided into $2^N$ contributions, where $N$ is the 
number of $i G_{1 1}$'s and $i G_{2 2}$'s. Above observation leads 
us to regard $i G_{1 2}$, $i G_{2 1}$, $i G_{1 1}^{(T)}$, and 
$i G_{2 2}^{(T)}$ as the {\em cutted propagators}. 

Through the applications of the above cutting rules, $A$ is divided 
into several subparts. Each subpart contains only type-1 vertices or 
only type-2 vertices. The former (latter) belongs to $S$ ($S^*$) in 
$\langle S^* S \rangle$. The cuttings work as follows. The line that 
cut $i G_{1 2} (P)$ with $p_0 > 0$ ($p_0 < 0$) is the initial-state 
particle (final-state antiparticle) cut line. The line that cut 
$i G_{2 1} (P)$ with $p_0 > 0$ ($p_0 < 0$) is the final-state 
particle (initial-state antiparticle) cut line. The line that cut 
$i G_{1 1}^{(T)} (P)$ [$i G_{2 2}^{(T)} (P)$] is the initial-state 
cut line {\em and} the final-state cut line in $S$ [$S^*$] and, in 
$S^*$ [$S$], an one extra spectator particle with $P$ is. For the 
line that cut $i G_{1 1}^{(T)} (P)$ with $p_0 > 0$ ($p_0 < 0$) is 
the initial-state particle (antiparticle) cut line {\em and} the 
final-state particle (antiparticle) cut line. For the cut line on 
$i G_{2 2}^{(T)} (P)$, similar statement holds. 

It is quite obvious that the \lq\lq cutting rules'' introduced above 
for thermal reaction rates may be used for general thermal 
amplitudes evaluated in the Keldish variant of RTF. 

Finally, it is worth mentioning that it can easily be seen from 
Eqs.~(\ref{rate}) and (\ref{cut-t}) that the RHS of 
Eq.~(\ref{cut-t}), which represents the imaginary part of a physical 
amplitude, is a sum of various reaction rates times corresponding 
kinematical factors. 
\section*{Acknowledgments}
This work was supported in part by the Grant-in-Aide for Scientific 
Research ((A)(1) (No.~08304024)) of the Ministry of Education, 
Science and Culture of Japan. 
\setcounter{equation}{0}
\setcounter{section}{1}
\section*{Appendix A Proof of Equation (3.24)} 
\def\theequation{\mbox{\Alph{section}.\arabic{equation}}}
Here we prove the identity Eq.~(\ref{koutou0}). We expand the 
RHS of Eq.~(\ref{koutou0}) in powers of $n_B (x) (\equiv \xi)$ to 
obtain 
\begin{eqnarray} 
& & \sum_{k = 0}^{\mbox{\scriptsize{min}} (j, \, j')} \frac{i !}{(i 
- j + k) !} \frac{i' ! \, j' !}{(j' - k) !} \left( 
\begin{array}{c} 
j \\ 
k 
\end{array} 
\right) \xi^{i + k} (1 + \xi)^{j' - k} \nonumber \\ 
& & \mbox{\hspace*{5ex}} = \sum_{k = 0}^{\mbox{\scriptsize{min}} 
(j, \, j')} \sum_{\ell = 0}^{j' - k} \frac{i !}{(i - j + k) !} \, 
\frac{i' !}{\ell ! \, (j' - k - \ell) !} \, \frac{j ! \, j' !}{k ! 
\, (j - k) !} \, \xi^{i + j' - \ell} \nonumber \\ 
& & \mbox{\hspace*{5ex}} = \sum_{k = 0}^{j'} \sum_{\ell = 
0}^{\mbox{\scriptsize{min}} (j' - k, \, j)} \frac{i !}{(i - j + 
\ell) !} \, \frac{j !}{\ell ! \, (j - \ell) !} \, \frac{i' ! \, j' 
!}{k ! \, (j' - k - \ell) !} \xi^{j + i' - k} \, , 
\label{A.1} 
\end{eqnarray} 

\noindent where $i \geq j$. Comparing Eq.~(\ref{A.1}) with 
Eq.~(\ref{first}), we see that it is sufficient to show that 
\begin{equation} 
\displaystyle{ 
\raisebox{0.9ex}{\scriptsize{$k$}}} \mbox{\hspace{-0.1ex}} 
{\cal F}^{i, \, i'}_{j, \, j'} = 
\displaystyle{ 
\raisebox{0.9ex}{\scriptsize{$k$}}} \mbox{\hspace{-0.1ex}} 
{\cal G}^{i, \, i'}_{j, \, j'} \, , 
\label{arama} 
\end{equation} 
where 
\begin{eqnarray} 
\displaystyle{ \raisebox{0.9ex}{\scriptsize{$k$}}} 
\mbox{\hspace{-0.1ex}} 
{\cal F}^{i, \, i'}_{j, \, j'} & \equiv & \sum_{\ell = 
0}^{\mbox{\scriptsize{min}} (j, \, j' - k)} 
\frac{i! \, j!}{\ell ! \, (i - 
j + \ell) ! \, (j' - k - \ell) ! \, (j - \ell) !} \, , 
\label{F} \\ 
\displaystyle{ \raisebox{0.9ex}{\scriptsize{$k$}}} 
\mbox{\hspace{-0.1ex}} {\cal G}^{i, \, i'}_{j, \, j'} & \equiv & 
\frac{(j + i' - k) !}{(i' - k) ! \, (j' - k) !} \, . 
\label{G} 
\end{eqnarray} 
Here we define two functions, 
\begin{eqnarray} 
F^{i, \, i'}_{j, \, j'} (x) & \equiv & \sum_{k = 0}^{j'} x^{j' - k} 
\, \, \displaystyle{ 
\raisebox{0.9ex}{\scriptsize{$k$}}} \mbox{\hspace{-0.1ex}} 
{\cal F}^{i, \, i'}_{j, \, j'} 
\label{F-func} \\ 
G^{i, \, i'}_{j, \, j'} (x) & \equiv & \sum_{k = 0}^{j'} x^{j' - k} 
\, \, \displaystyle{ 
\raisebox{0.9ex}{\scriptsize{$k$}}} \mbox{\hspace{-0.1ex}} 
{\cal G}^{i, \, i'}_{j, \, j'} \, . 
\label{G-func} 
\end{eqnarray} 
It can easily be shown that $F$'s and $G$'s satisfy the same 
differential equation, 
\begin{eqnarray} 
\frac{d}{d x} F^{i, \, i'}_{j, \, j'} (x) & = & F^{i, \, i' - 
1}_{j, \, j' - 1} (x) + 
j \, F^{i, \, i'}_{j - 1, \, j' - 1} (x) \, , 
\label{F-diff} \\ 
\frac{d}{d x} G^{i, \, i'}_{j, \, j'} (x) & = & G^{i, \, i' - 
1}_{j, \, j' - 1} (x) + 
j \, G^{i, \, i'}_{j - 1, \, j' - 1} (x) \, . 
\label{G-diff} 
\end{eqnarray} 
From Eqs.~(\ref{F-func}), (\ref{G-func}) with Eqs.~(\ref{F}) and 
(\ref{G}), we obtain 
\begin{eqnarray} 
F^{i, \, i'}_{j, \, j'} ( 0 ) & = & G^{i, \, i'}_{j, \, j'} ( 0 ) 
= \frac{i !}{(i - j) !} 
\label{ini} \\ 
F^{i, \, i'}_{j, \, 0} ( x ) & = & G^{i, \, i'}_{j, \, 0} ( x ) 
= \frac{i !}{(i - j) !} \, . 
\label{bound} 
\end{eqnarray} 

We see from Eq.~(\ref{F-diff}) [Eq.~(\ref{G-diff})] that 
$F^{i, \, i'}_{j, \, j'} (x)$ [$G^{i, \, i'}_{j, \, j'} (x)$] may be 
obtained from $F^{i, \, \hat{i}'}_{\hat{j}, \, 0} (x)$ [$G^{i, \, 
\hat{i}'}_{\hat{j}, \, 0} (x)$] in Eq.~(\ref{bound}) with $\hat{i}' 
\leq i'$, $\hat{j} \leq j$, and $F^{i, \, \hat{i}'}_{\hat{j}, \, 
\hat{j}'} (0)$ [$G^{i, \, \hat{i}'}_{\hat{j}, \, \hat{j}'} (0)$] in 
Eq.~(\ref{ini}) with $\hat{i}' \leq i'$, $\hat{j} \leq j$, $\hat{j}' 
\leq j'$. Since $F$'s and $G$'s subject to the same set of equations 
(\ref{F-diff}) - (\ref{bound}), we conclude that  
\[ 
F^{i, \, i'}_{j, \, j'} (x) = G^{i, \, i'}_{j, \, j'} (x) \, , 
\] 
which proves Eq.~(\ref{arama}). 
\begin{flushright} 
Q.E.D. 
\end{flushright} 

\newpage 
\begin{description} 
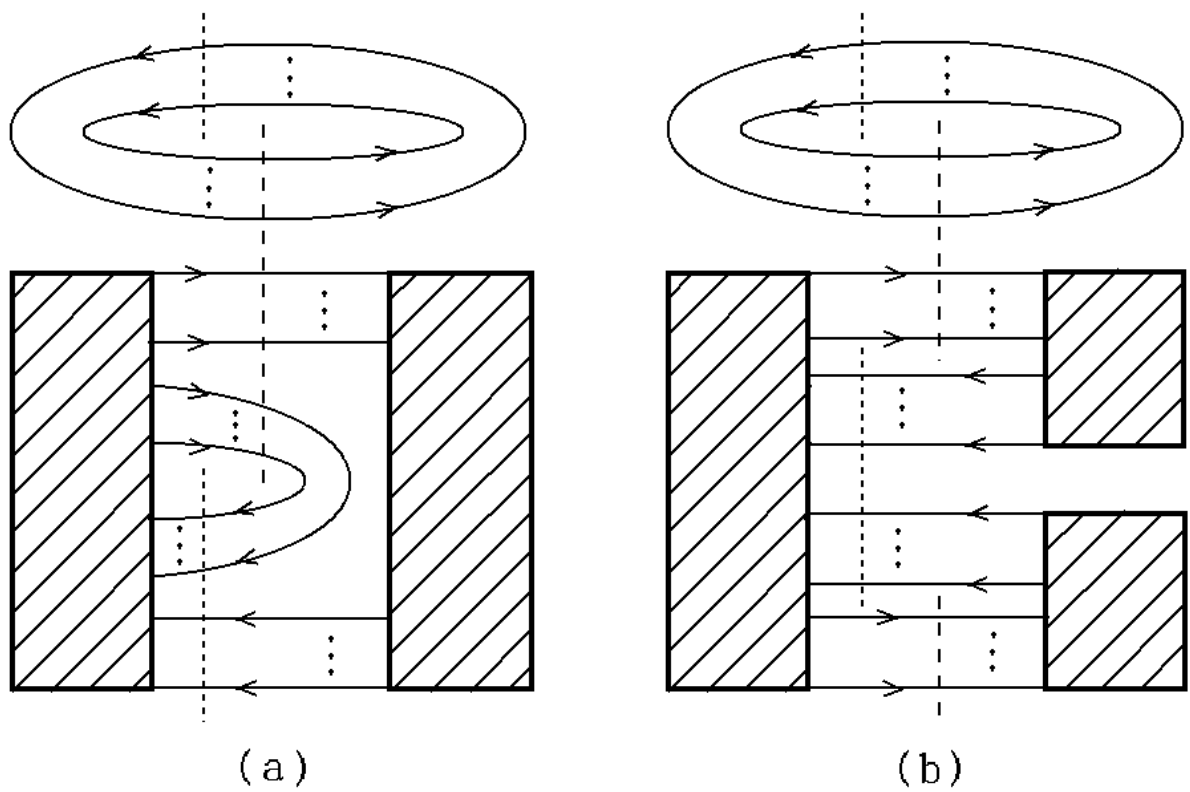
\item{FIG. 1.} 
Two examples of double-cut diagrams for the transition 
probability $W = S^* S$ in vacuum theory. Dashed lines are the 
final-state cut lines while the dotted lines are the initial-state 
cut lines. The left side of the cut lines represents the $S$-matrix 
element, $S$, while the right side does $S^*$. The line that is 
cutted by the final-state (initial-state) cut line represents a 
particle in the final (initial) state in $S$. The lines cutted by 
the initial-state [final-state] cut line include those corresponding 
to $\{ A \}$ [$\{ B \}$] in Eq.~(\ref{2}). The group of lines on top 
of diagrams stands for spectator particles. (a) Both $S$ and $S^*$ 
are connected. In addition to the spectator particles mentioned 
above, additional spectator particles are in $S^*$. (b) $S$ is 
connected while $S^*$ is disconnected. Note, however, that $S^* S$ 
is connected. 
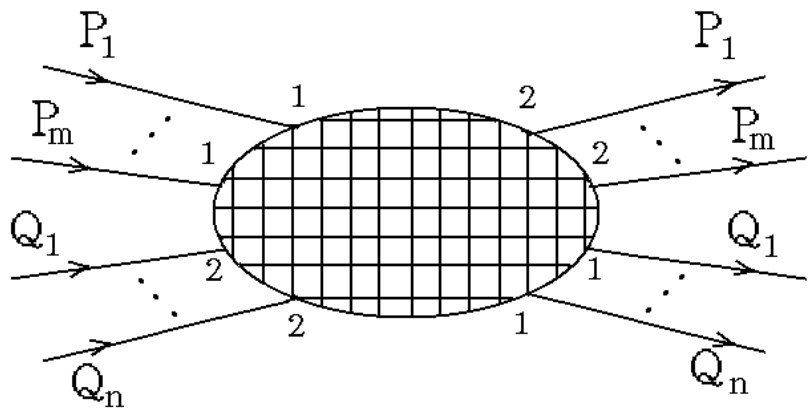
\item{Fig. 2} Diagrammatic  representation of the thermal amplitude 
$A$ in Eq.~(\ref{rate}). 
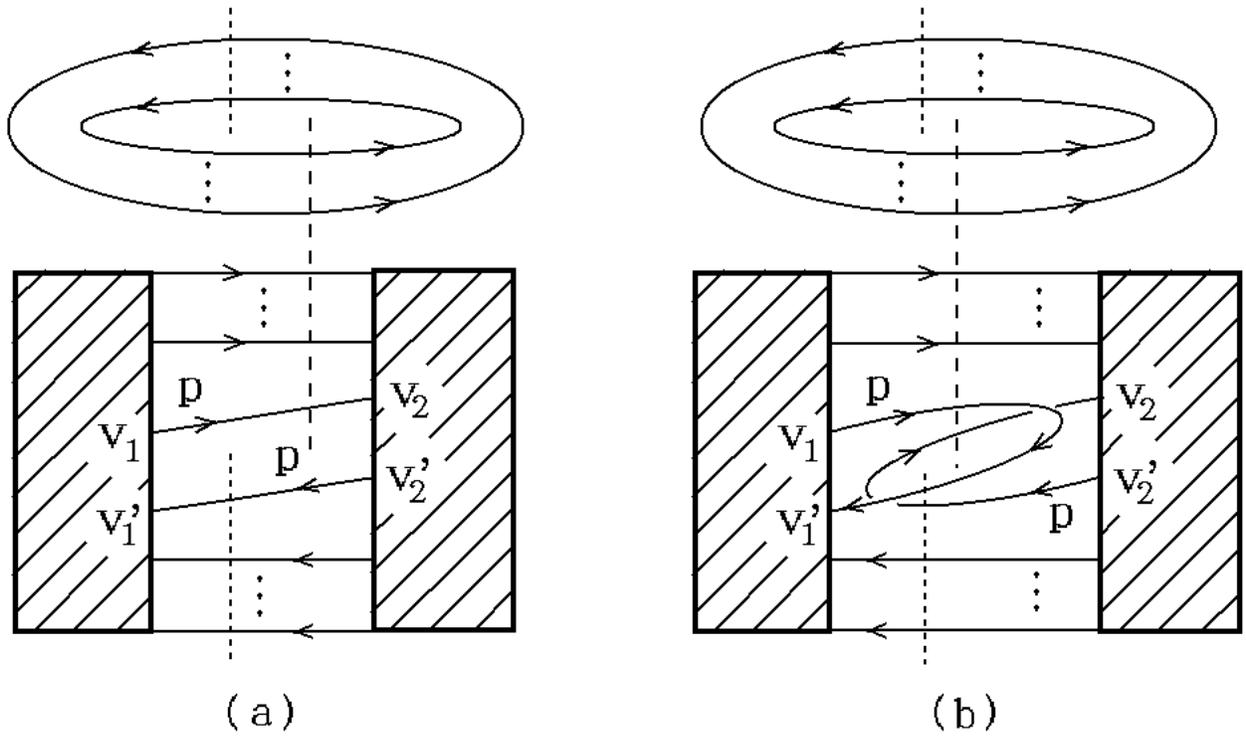
\item{Fig. 3} Double-cut diagrams for $W = S^* S$, which yields (a) 
$i D^{(+)}_{1 2} (P) \, i D^{(+)}_{2 1} (P)$ and (b) $i D^{(T) \, 
(+)}_{1 1} (P) \, i D^{(T) \, (+)}_{2 2} (P)$. Here $P = (p, 
{\bf p})$. 
\item{Fig. 4} Double-cut diagrams for $W = S^* S$, which yields (a) 
$i D^{(+)}_{1 2} (P) \, i D^{(+)}_{2 1} (P)$ and (b) $i D^{(T) \, 
(+)}_{1 1} (P) i D^{(T) \, (+)}_{2 2} (P)$. Here $P = (p, {\bf p})$. 
\end{description} 
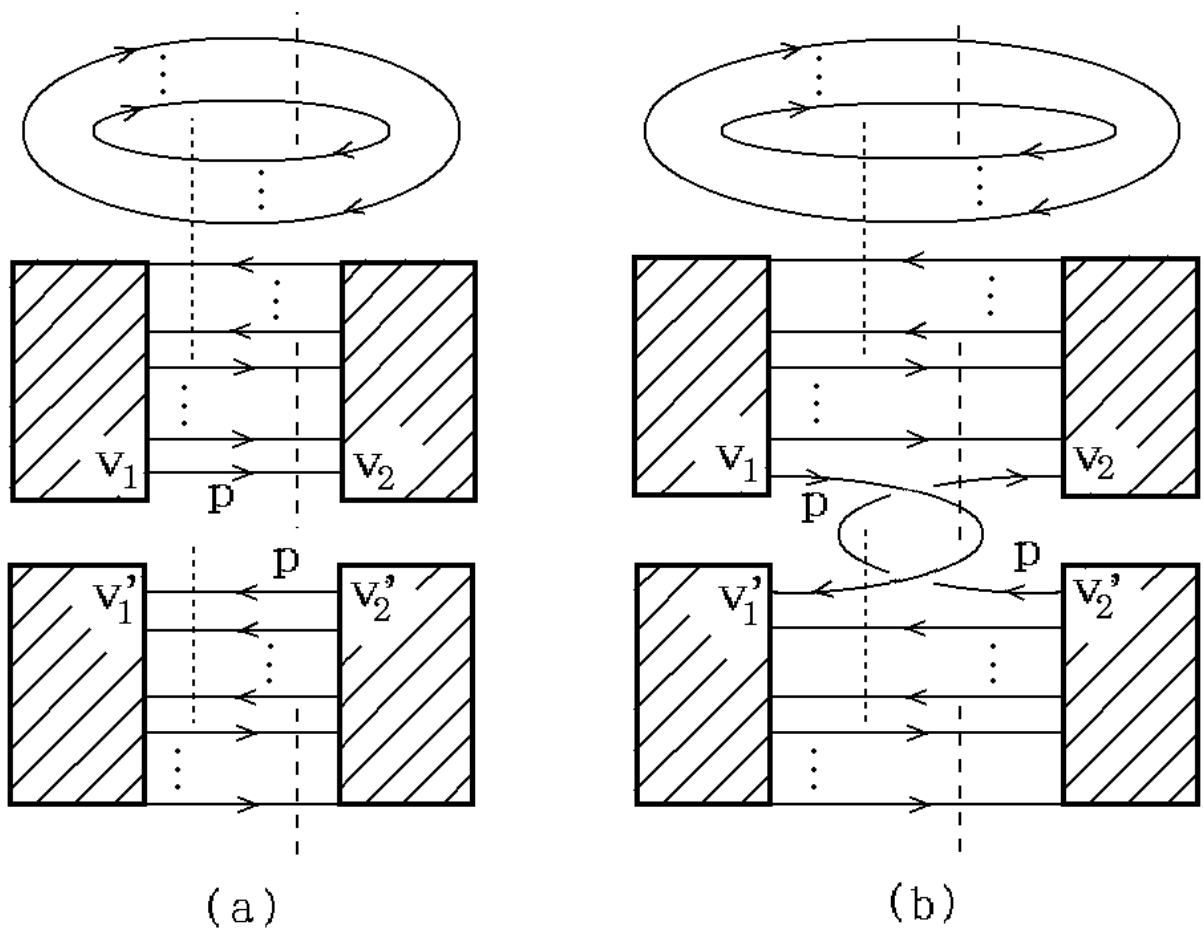
\end{document}